\documentclass{aa}  

\usepackage[colorlinks]{hyperref} 
\usepackage{graphicx}
\usepackage{txfonts}
\begin{document}

   \title{Mapping the working of environmental effects in A963}

   \subtitle{}

   \author{Boris Deshev\inst{1,2}
          \and Christopher Haines\inst{3}
          \and Ho Seong Hwang\inst{4}
          \and Alexis Finoguenov\inst{5}
          \and Rhys Taylor\inst{1}
          \and Ivana Orlitova\inst{1}
          \and Maret Einasto\inst{2}
		  \and Bodo Ziegler\inst{6}
          }

   \institute{Astronomical Institute, Czech Academy of Sciences, Bo\u{c}n\'i II 1401, CZ-14131 Prague, Czech Republic\\
              \email{deshev@asu.cas.cz}
         \and
				University of Tartu, Tartu Observatory, Estonia         
		 \and
				Instituto de Astronom\'ia y Ciencias Planetarias de Atacama, Universidad de Atacama, Copayapu 485, Copiap\'o, Chile
		 \and
		 		Korea Astronomy and Space Science Institute, 776 Daedeokdae-ro, Yuseong-gu, Daejeon 34055, Korea
		\and
                Department of Physics, University of Helsinki, PO Box 64, FI-00014 Helsinki, Finland
		 \and
		 		Institute for Astrophysics, University of Vienna, 1180 Wien, Austria
                        }         

   \date{}

 
  \abstract
   {}
   {We qualitatively assess and map the relative contribution of pre-processing and cluster related processes to the build-up of A963, a massive cluster at z = 0.2 showing an unusually high fraction of star forming galaxies in its interior.}
   {We use Voronoi binning of positions of cluster members on the plane of the sky in order to map the 2D variations of galaxy properties in the centre and infall region of A963. We map four galaxy parameters (fraction of star forming galaxies, specific star formation rate, {\sc Hi} deficiency and age of the stellar population) based on full SED fitting, 21 cm imaging and optical spectroscopy.}
   {We find an extended region dominated by passive galaxies along a north--south axis crossing the cluster centre, possibly associated with known filaments of the large-scale structure. There are signs that the passive galaxies in this region were quenched long before their arrival in the vicinity of the cluster. Contrary to that, to the east and west of the cluster centre lie regions of recent accretion dominated by gas rich, actively star forming galaxies not associated with any substructure or filament. The few passive galaxies in this region appear to be recently quenched, and some gas rich galaxies show signs of ongoing ram-pressure stripping. We report the first tentative observations at 21 cm of ongoing ram-pressure stripping at $z$ = 0.2, as well as observed inflow of low-entropy gas into the cluster along filaments of the large-scale structure.}
   {The observed galaxy content of A963 is a result of strongly anisotropic accretion of galaxies with different properties. Gas rich, star forming galaxies are being accreted from the east and west of the cluster and these galaxies are being quenched at r < R$_{200}$, likely by ram-pressure stripping. The bulk of the accretion onto the cluster, containing multiple groups, happens along the north--south axis and brings mostly passive galaxies, likely quenched before entering A963.}


   \maketitle
%

\section{Introduction}

The physics driving the empirically established dependence of most galaxy properties on their environment is still poorly constrained \citep{Boselli&Gavazzi2006, ParkHwang2009, BoselliGavazzi2014}. Ram-pressure stripping \citep{Gunn&Gott1972, Quilis2000, Vollmer2001, Bekki&Couch2003} is often invoked to explain the observations of gas-deficient galaxies or galaxies with irregular appearance in clusters \citep{KenneyvanGorkom&Vollmer2004, Chung2007, Poggianti2017, Jachym2019, Ebeling2019}. It has also been successfully shown to explain the distribution, in projected phase-space, of passive galaxies in clusters \citep{Jaffe2016, Yoon2017, Jaffe2018}. However, because its strength depends on galaxies moving at high relative velocities through a relatively dense gas, ram-pressure stripping is only expected to be efficient in massive clusters of galaxies. In reality the distribution of galaxy properties throughout the large-scale structure of the universe requires a complex combination of multiple physical mechanisms. Some of the galaxies are quenched well before they find their way to the interiors of clusters \citep{Bianconi2018}. At low redshifts, approximately 40\% of the galaxies residing outside clusters show no ongoing star formation \citep{Balogh1999,Verdugo2008}, a fact which cannot be explained with ram-pressure stripping. A collective term, pre-processing, is often used to explain the presence of this passive population and includes a number of physical mechanisms acting on galaxies residing in smaller haloes or filaments. Most likely starvation is the mechanism responsible for this \citep{Larson1980, Balogh2000, Bekki2002, Goto2003} as it is capable of affecting a large fraction of all the galaxies, the ones that are satellites in a larger halo but not necessarily in a cluster of galaxies. Starvation is the absence of accretion of fresh gas onto galaxies that can be used for star formation. It has been shown that the star formation histories of the passive galaxies in the local universe are consistent with starvation being the main mechanism responsible for quenching their star formation \citep{Peng2015}. \citet{Einasto2014} showed that the fraction of passive galaxies in groups increases with group richness, and that the environment in which the groups reside modulates this trend. There are also other mechanisms that can alter the star formation properties of galaxies in lower density environments \citep{Cowie&Songaila1977, Fujita2004}. Thus, galaxies that have been pre-processed undergo little change when they reach the interiors of massive clusters. Other galaxies, likely individual, only stop forming stars when they cross the virial radius of a cluster and are observed as jellyfish galaxies \citep{Poggianti2017}. In addition, high-density environments, be it groups, clusters, or non-virialised structures like filaments, are associated not only with quenching, but sometimes also with an increase in star formation \citep{Fujita&Nagashima1999, HwangLee2009, Stroe2015a, Deshev2017, Vulcani2019}.

Abell 963 (hereafter A963) is one of the best studied clusters outside the local universe. It has been targeted with XMM-Newton, Galex, Spitzer, and the LMT \citep{Cybulski2016}. The cluster has been observed in the optical with the Isaac Newton Telescope, and virtually all of its members brighter than R$\leq$19 have been targeted for optical spectroscopy \citet[hereafter J16]{Jaffe2016}. A963 is one of only two clusters outside the local universe for which we have 21cm data, a result of a blind interferometric survey with the WSRT called BUD{\sc Hi}ES \citep[][Gogate et al. \textit{MNRAS accepted}]{Verheijen2007, Deshev2009}. The cluster and its constituent galaxies have been extensively studied \citep[J16]{Jaffe2013}. The projected phase-space analysis of A963 presented in J16 explains to a remarkable extent the distribution of the galaxies with different properties residing in this cluster, and confirms the important role of ram-pressure stripping as the reason for the high passive fraction of cluster galaxies. However, the projected phase-space analysis, particularly when applied to individual clusters, ignores part of the information contained in the data, like possible deviations from spherical symmetry of the cluster and anisotropic accretion of galaxies. Although some clusters might indeed be close to spherical, this is certainly not the case for clusters experiencing recent mergers or significant accretion. The accretion of galaxies onto clusters is likely not isotropic both in terms of the number of galaxies falling from different directions and the level of pre-processing of these galaxies \citep{Martinez2016, Kuutma2017, Odekon2018}. Observations and simulations show that 35\%--40\% of galaxies, by luminosity, are found in filaments \citep{Tempel2014}. In addition, the galaxies accreted along filaments are subjected to pre-processing \citep{Kuutma2017,Vulcani2019}, so they have different properties from those accreted individually. 

In this article we concentrate on the distribution of galaxies with different properties on the sky, complementing the analysis presented in J16. While this falls short of the desired full, observed, phase-space analysis, we demonstrate that it is very informative about the fate of galaxies infalling onto large clusters, and the extent to which some of them are pre-processed. We concentrate on the galaxy members of A963\_1 , which is the main cluster \citep{Jaffe2013}, and we refer to it as A963 for the remainder of this article.

Throughout, we use a standard $\Lambda$CDM cosmology with H$_\mathrm{0}=70$ km s$^{-1}$ Mpc$^{-1}$, $\Omega_\mathrm{m}=0.3$, and $\Omega_\mathrm{\Lambda}=0.7$. A963 is observed at z = 0.205 ($\sim$ 1.0 Gpc), thus 1\arcsec corresponds to 3.36 kpc.

\section{Data}\label{data_section}
We combine a multi-fibre optical spectroscopy of galaxies, with rest-frame 21 cm imaging (WSRT), and XMM-Newton X-ray imaging. All the data used for the analysis presented in this article have been used in previous publications or have been publicly released. Here we briefly describe the data while presenting the measurements of the individual galaxy properties. The lead data set for this study is a combination of two separate optical spectroscopy runs of galaxies in the field of A963, both executed with Hectospec on the MMT\footnote{Observations reported here were obtained at the MMT Observatory, a joint facility of the University of Arizona and the Smithsonian Institution.}. Together they contain optical spectra of 1712 galaxies within the square degree surrounding the cluster centre. Approximately 3/4 of these data come from the ACReS survey \citep{Haines2013}, which is the spectroscopic part of the LoCuSS survey \citep{Smith2010a}. The rest are from a survey published by \citet{Hwang2014}. There are many galaxies targeted by both surveys. In this study we select all the targets from ACReS and add the unique data from \citet{Hwang2014}. These data formed the majority of the data set (>93\%) analysed by J16.

\subsection{Cluster membership}\label{cluster_membership}
\begin{figure}
   \centering
   \includegraphics[width=\hsize]{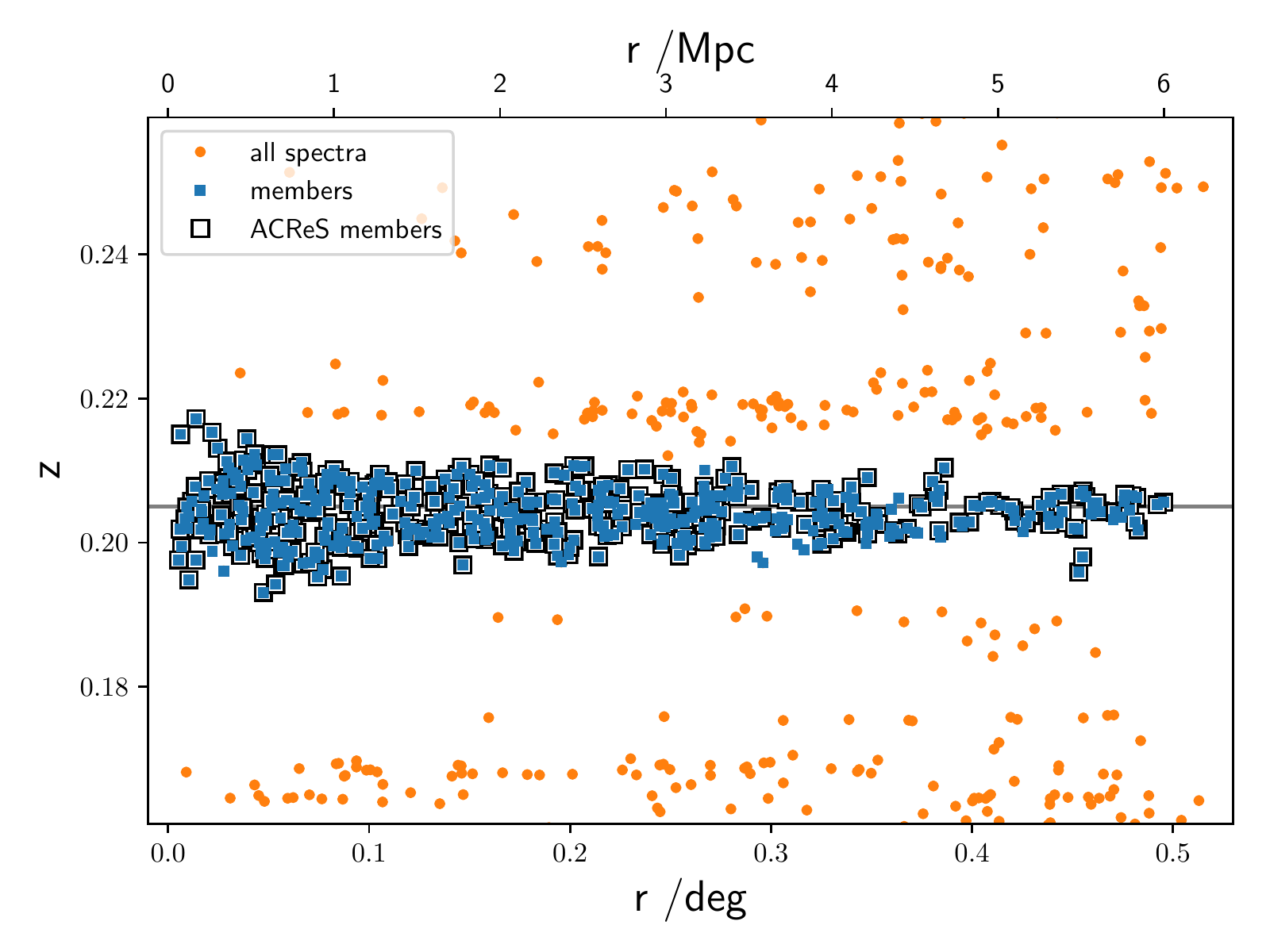}
      \caption{Selection of the cluster members (blue squares) of A963 in projected phase space, redshift vs. cluster-centric distance. The cluster members from the ACReS survey have black squares around their symbols. The orange dots show all the galaxies with spectroscopy in the field. The redshift of the cluster (z = 0.205) is shown with the grey horizontal line.}
         \label{memb_selection}
   \end{figure}

Since most of our data come from the ACReS survey ($\sim$ 75\%), we adopted their selection of cluster members \citep{Haines2015} based on the distribution of galaxies in projected phase space. Because of the shape of the gravitational potential of the cluster, the member galaxies form a well-defined trumpet shape. While adding the additional data we overplotted them onto the ACReS cluster members, and selected the galaxies overlapping with the already defined cluster region. This process is shown in Fig.\ref{memb_selection}, where the 481 cluster members are shown with blue squares, while those from the ACReS survey are outlined with black squares. The distribution of the remaining galaxies with available spectroscopy (orange dots) clearly shows the presence of other structures around the main cluster, as discussed in \cite{Jaffe2011}. The central redshift of the cluster, (z = 0.205) measured by \cite{Deshev2017} is shown with horizontal grey line on the figure. We use the position of the brightest cluster galaxy as a cluster centre (R.A.=$10^{\mathrm{h}} 17^{\mathrm{m}} 03\fs60$, Dec=$+39\degr 02\arcmin 49\farcs920$), measured from the R-band imaging of the cluster with INT, La Palma, collected as part of the BUD{\sc Hi}ES survey (Gogate et al. \textit{MNRAS accepted}). The X-ray emission does not show any significant displacement from this point on all scales. The final number of cluster members used in this analysis is 386, after applying a stellar mass cut and removing galaxies with active galactic nuclei (Section \ref{sect_smcgc}). The brightest cluster galaxy is not part of this spectroscopic survey.

\begin{figure}
   \centering
   \includegraphics[width=\hsize]{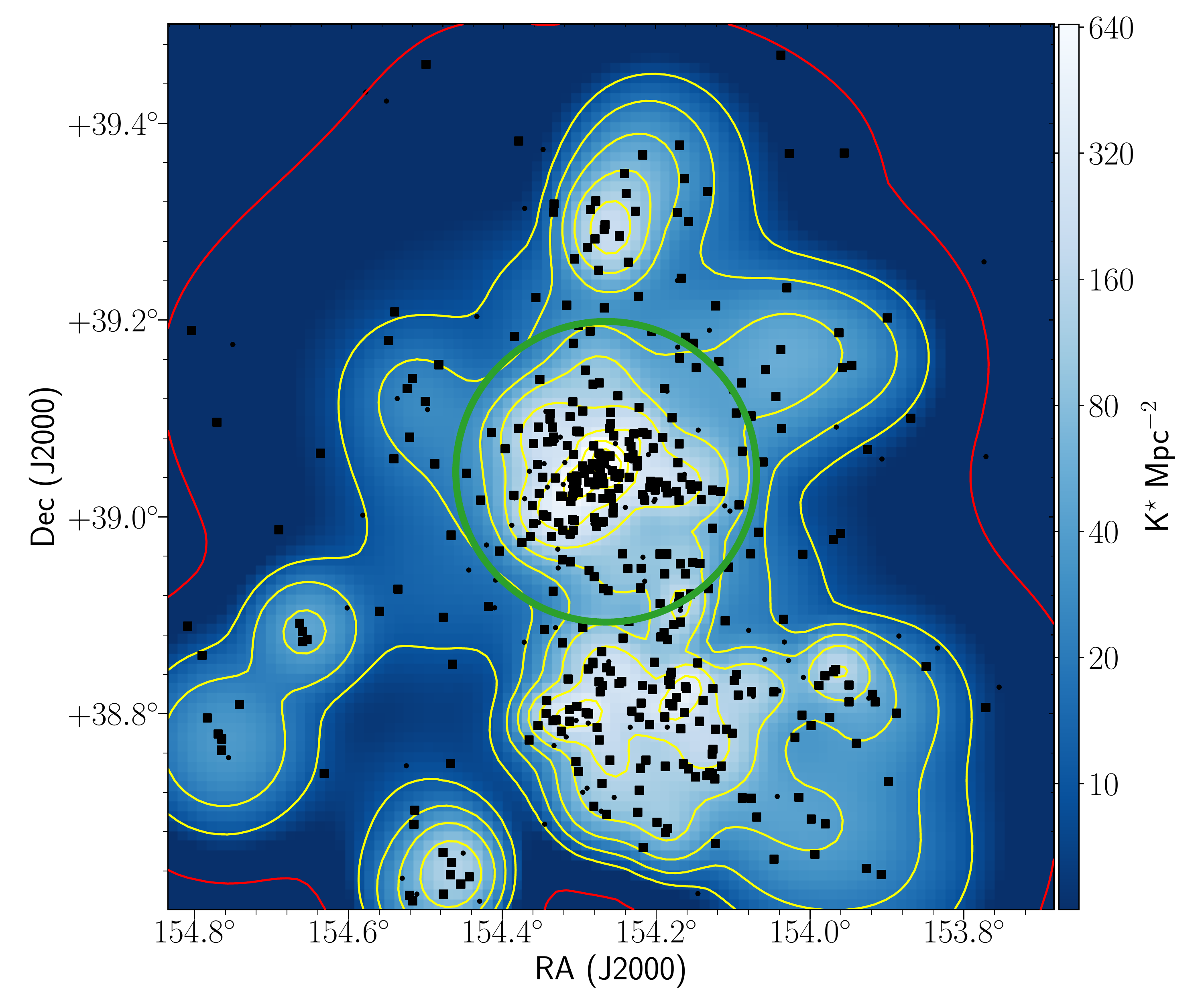}
      \caption{K-band luminosity density of the A963 cluster members. Solid squares show cluster members with log(M$_{\star})\geq$ 10.0 M$_{\odot}$. Small points show cluster members with lower stellar mass. The yellow contour levels are indicated by the tick marks on the colour bar. The green circle has a radius equal to R$_{200}$ of the cluster. The red contour, at 1.8 K$^{\star}$ Mpc$^{-2}$, is used to define the radial extent of the rest of the maps in the article.}
         \label{klum}
   \end{figure}

Figure \ref{klum} shows the K-band luminosity density based on all the cluster members. The colour map, together with the logarithmically spaced yellow contours, show the main overdensity of A963 together with a number of other structures clustered predominantly along the north--south axis of the cluster. The red contour is drawn at 1.8 K$^{\star}$ Mpc$^{-2}$. This level was chosen because it encompasses all the cluster members and is used, together with a circle with a radius of 0.52\degr, to define the radial extent of the maps of the cluster presented in the following sections.
   \begin{figure}
   \centering
   \includegraphics[width=\hsize]{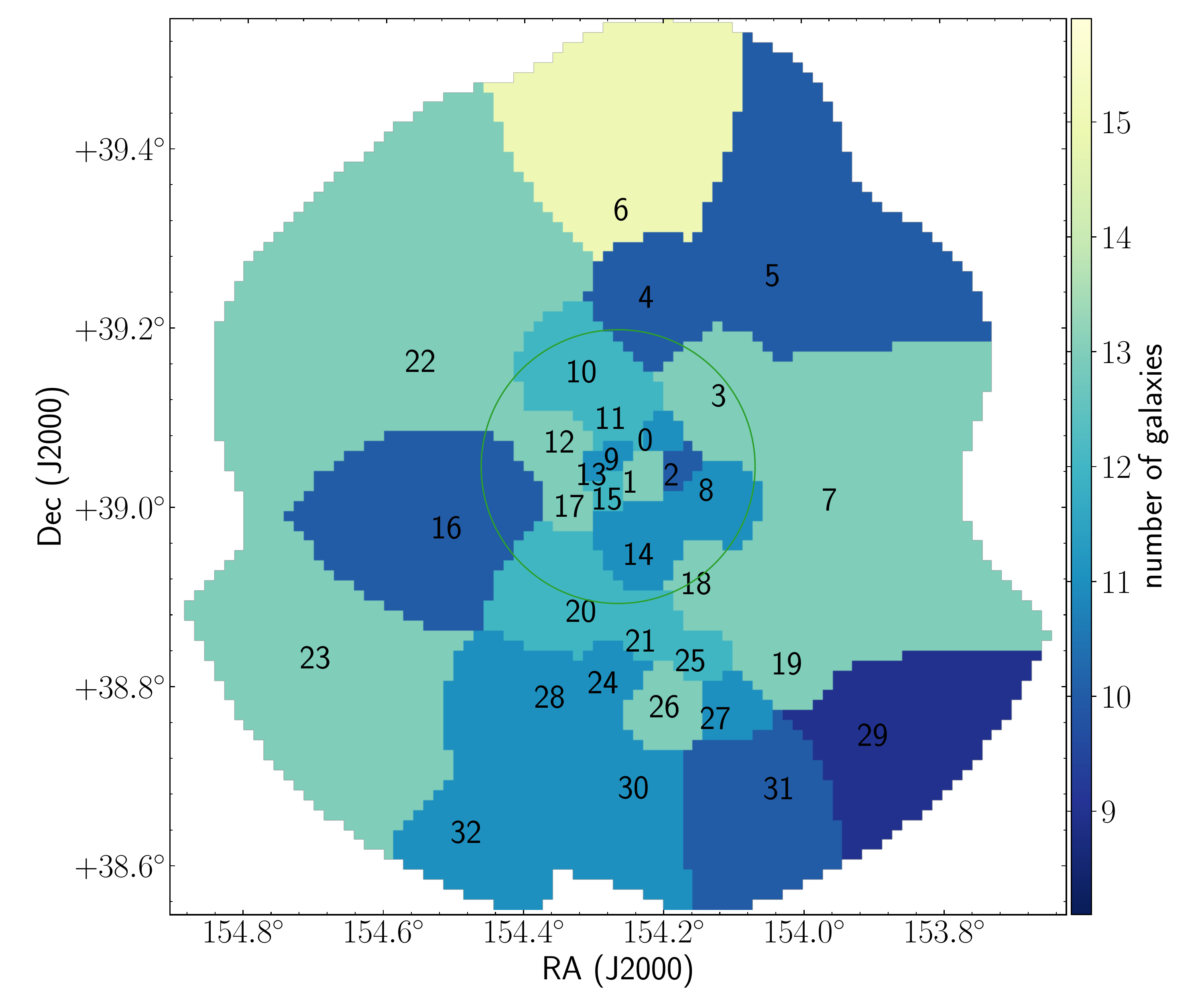}
      \caption{Map of the individual Voronoi bins. Each bin is coloured according to the total number of galaxies within it. All the bins are numbered, starting with zero. The green circle has a radius equal to R$_{200}$ of the cluster.}
         \label{bin_map}
   \end{figure}
   \begin{figure}
   \centering
   \includegraphics[width=\hsize]{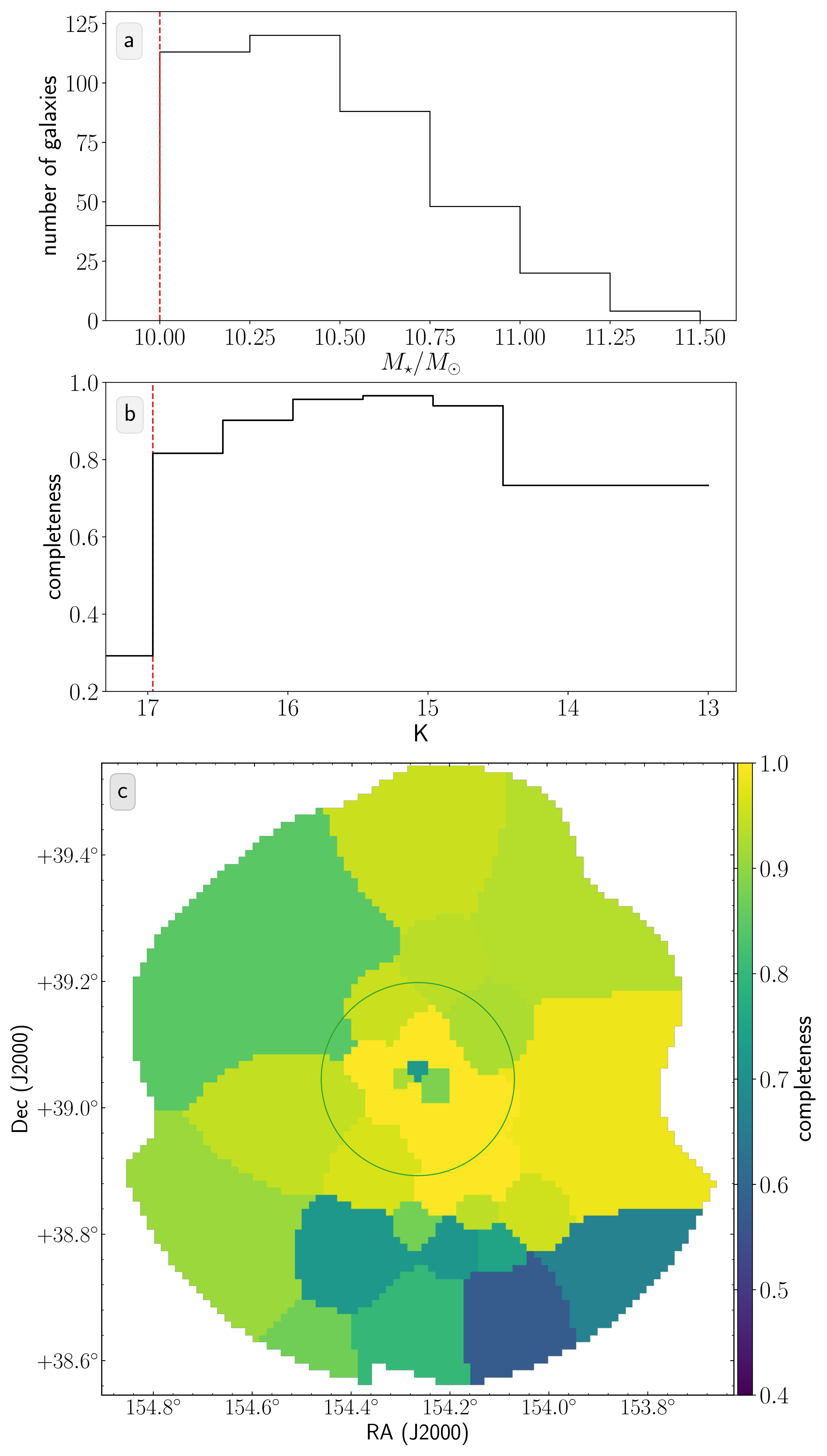}
      \caption{Panel a: Stellar mass distribution of the cluster members with the completeness limit indicated by the red, dashed line. Panel b: Completeness of our data set with respect to K-band luminosity. The red dashed line is at K = K$^{\star}$+2.0- the magnitude limit of ACReS survey. Panel c: Completeness distribution on the sky. The green circle has a radius equal to R$_{200}$ of the cluster.}
         \label{completeness}
   \end{figure}
   \begin{figure}
   \centering
   \includegraphics[width=\hsize]{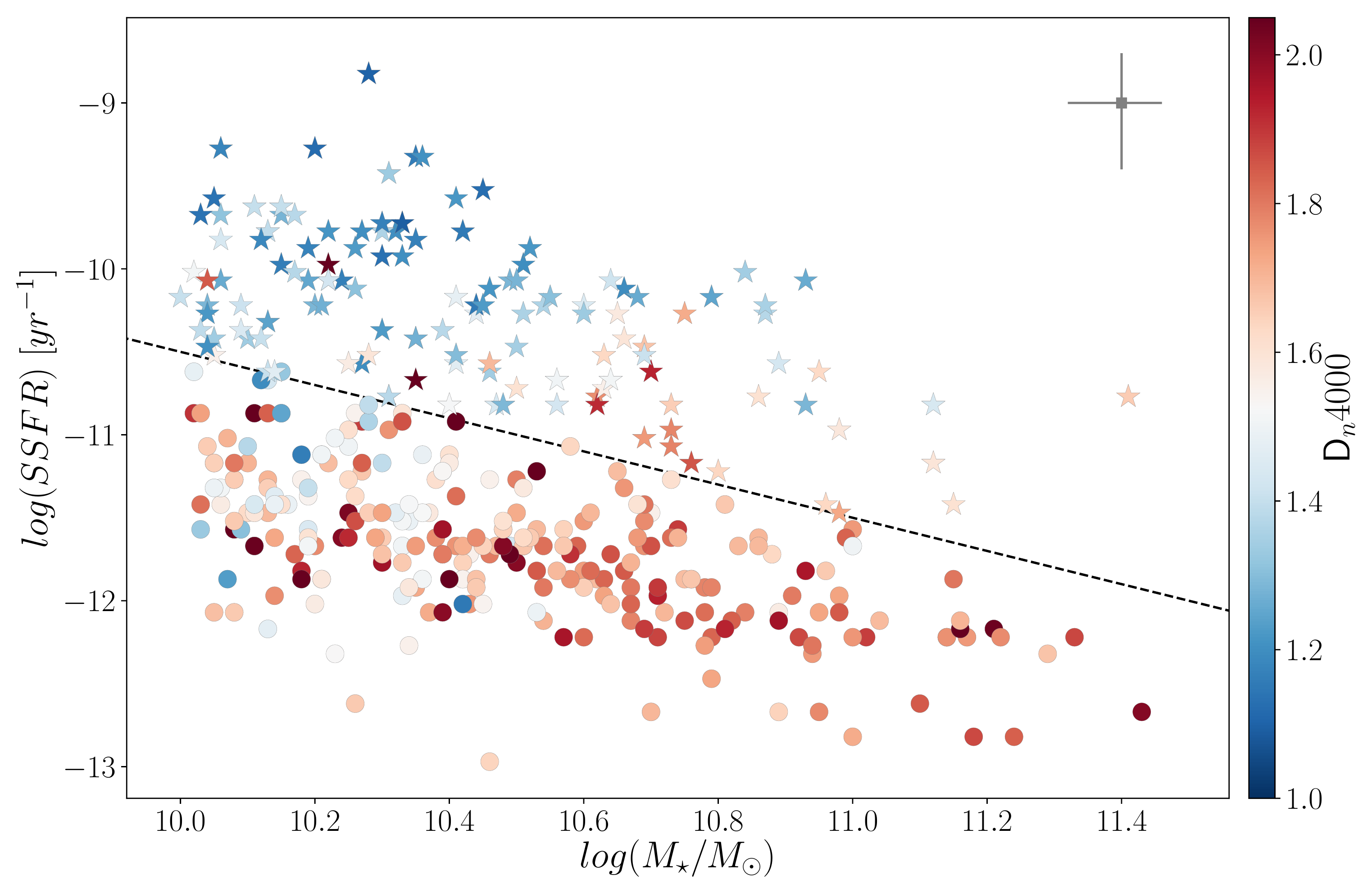}
      \caption{Specific star formation rate of the A963 member galaxies as a function of their stellar mass. The black, dashed line at constant log(SFR) = -0.5 is used to split the galaxies into star forming, shown with stars, and passive, shown with circles. Symbols are coloured according to the strength of the 4000\AA~drop in luminosity. The symbol in the upper right corner shows median uncertainties.}
         \label{ssfr}
   \end{figure}

\subsection{Voronoi binning}\label{vorbin_section}
Binning of data with variable bin size is a common way of achieving constant signal-to-noise ratio across the bins. One way of calculating adaptive bins on a 2D plane is Voronoi binning, often used with integral-field unit (IFU) observations of galaxies, where the exponential drop in surface brightness requires binning away from the galactic centre. Mapping the properties of galaxies residing in clusters presents similar problems due to the drop in number density away from the cluster centre. We use Voronoi binning, as implemented by \citet{Cappellari&Copin2003} in the python procedure \textit{Vorbin}. The program is designed to work with IFU data, and as such it requires the data to be on a regular grid with well-defined pixels. We start by putting a square sky area centered on A963 on a regular grid with 90 $\times$ 90 cells and with 1.1\degr on the side. This number of cells is large enough so that the most populated one contains only five galaxies. At the same time it is small enough to guarantee the fast working of \textit{Vorbin}, as the computational time required scales linearly with the number of pixels to bin. In addition, we changed the \textit{Vorbin} subroutine calculating the signal-to-noise ratio in the bins to a simple sum in order to achieve a semi-constant number of galaxies per Voronoi bin. We required the Voronoi bins to contain about ten galaxies each in order to get meaningful statistics over each bin. In reality, because the geometry of the final bins is also taken into account \citep{Cappellari&Copin2003} the final number of galaxies in each bin varies between 9 and 15 with a median of 12. The 386 galaxies are assigned to 33 Voronoi bins. A map of the Voronoi bins with their number, referred to in the rest of the article, and showing the number of galaxies in each bin is shown in Fig. \ref{bin_map}. The green circle at the centre of the figure has a radius equal to R$_{200}$ of the cluster as calculated by J16 (R$_{200}$ = 1.85 Mpc), based largely on the same data set. The map covers the area where the K-band luminosity density is higher than 1.8 K$^{\star}$ Mpc$^{-2}$.

\subsection{Stellar mass, spectroscopic completeness, and galaxy classification}\label{sect_smcgc}
The stellar mass and star formation rate of all the cluster members was calculated with Multi-wavelength Analysis of Galaxy Physical properties \citep[MAGPHYS,][]{daCunha2008}. The fit was based on 18 bands covering wavelengths from far-infrared to far-ultraviolet, assuming Chabrier IMF \citep[see][for details]{Bianconi2020}. The distribution of stellar mass of A963 cluster members is shown in panel (a) of Fig.\ref{completeness}. Our completeness limit of log(M$_{\star}$)$\geq$10.0 M$_{\odot}$ is shown with the vertical red, dashed line. The stellar mass complete data set comprises 393 galaxies, members of A963. The middle panel "b" of Fig.\ref{completeness} shows the distribution of K-band luminosities of  the cluster members. The red dashed line shows the magnitude limit for target selection of the ACReS survey K = 16.964, equal to K$^{\star}$ + 2.0. The data used in this work is 89\% complete for galaxies brighter than this limit.

The presence of AGN at the centre of a galaxy could produce a spectral signature similar to that of ongoing star formation. We use the so-called BPT diagram \citep{BPT1981} to detect emission line galaxies with ionisation state of their interstellar matter (ISM) consistent with ionisation from AGN. The method uses a combination of two line ratios, {\sc [O\,i\,i\,i]} 5007\AA/H$\beta$ and {\sc [N\,i\,i]} 6583\AA/H$\alpha$. A line separating the AGNs from star forming galaxies from \cite{Kewley2001} was used. In total we found seven AGNs among the cluster members. These galaxies were not considered in the following analysis.

In this work we are interested in the properties of galaxies in different parts of the cluster, as projected on the sky. In panel (c) of Fig.\ref{completeness} we show the completeness on the sky of the data set used in this work. The spectroscopic completeness is shown as a fraction of all the galaxies with  K $\leq$ 16.964 and J-K colour consistent with the main sequence of the cluster. The lowest spectroscopic completeness of our data set is 58\%. More than 84\% of the sky area covered by this survey has spectroscopic completeness greater than 0.8.

Figure \ref{ssfr} shows the specific star formation rate (SSFR, defined as the star formation rate per unit stellar mass) of the cluster members as a function of their stellar mass. The dashed, black line at constant log(SFR) = -0.5 is used to divide the galaxies into star forming and passive, shown with stars and circles, respectively. Each symbol is colour-coded according to that galaxy's D$_{n}4000$ value (see Section \ref{d4000_section} for details), which is a good indicator of the age of the light-dominating stellar population. The error bar in the corner shows the median 1$\sigma$ confidence limits on the stellar mass and specific star formation rate from the $\chi^2$ fitting with MAGPHYS.

   \begin{figure}
   \centering
   \includegraphics[width=\hsize]{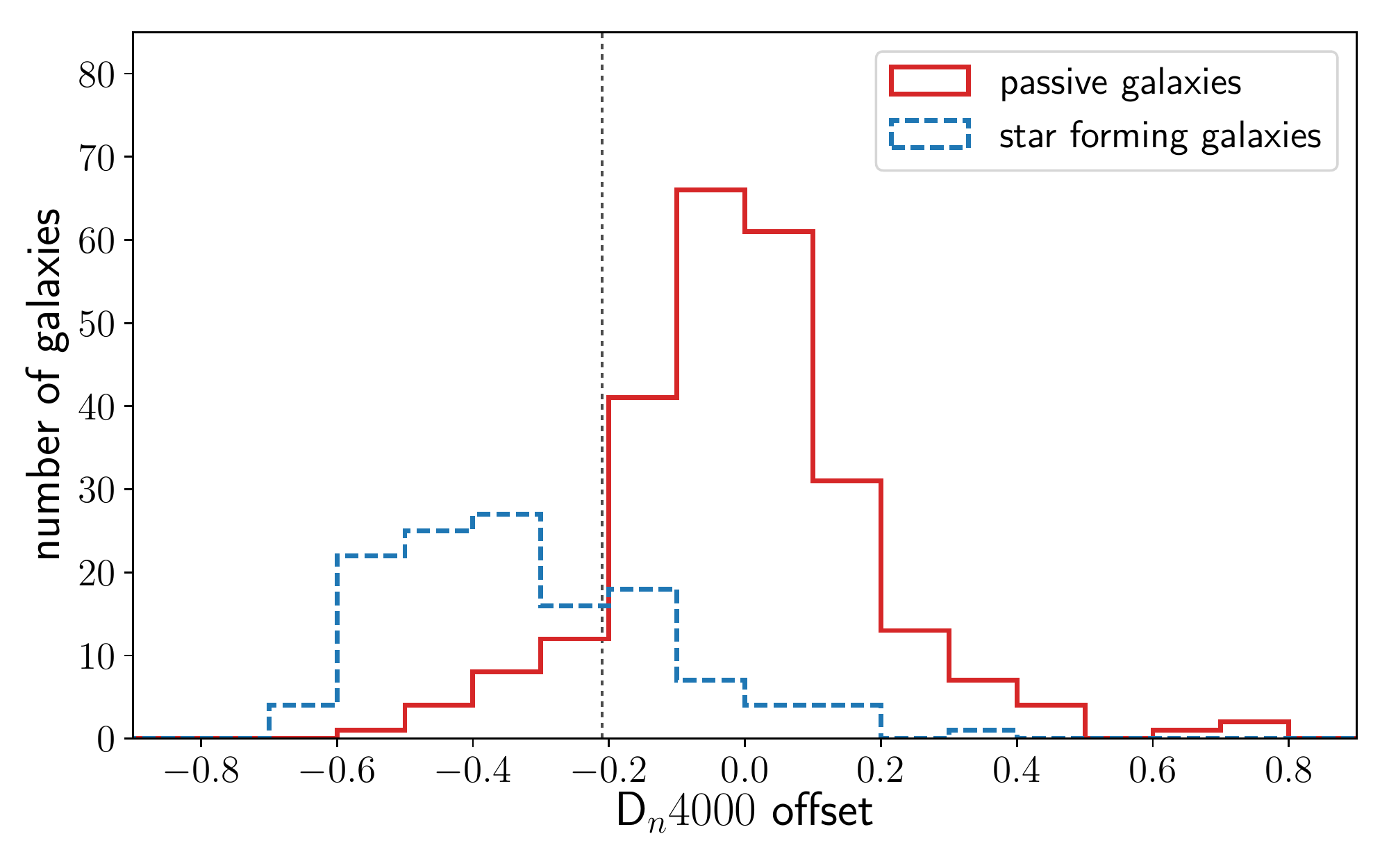}
      \caption{Distribution of D$_{n}4000$ values of A963 member galaxies. Plotted is the offset from the mean relation between stellar mass and D$_{n}4000$ for all the passive galaxies in the ACReS survey.}
         \label{D4000hist}
   \end{figure}

\subsubsection{D$_{n}4000$: Recently quenched and red star forming galaxies}\label{d4000_section}
Passive galaxies exhibit a characteristic break in their continuum emission blueward of 4000\AA~. This is due to accumulation of metal absorption lines coming from the atmospheres of low-mass stars and is known to correlate well with the age of the light-dominating stellar population \citep{Bruzual1983, PoggiantiBarbaro1997, Balogh1999}. We use the narrow definition of the index D$_{n}4000$ \citep{Balogh1999} for its robustness against reddening. The mean uncertainty of the individual D$_{n}4000$ measurements is 0.02. This number is confirmed by comparing the measurements of D$_{n}4000$ of galaxies for which more than one observation is available. In these cases the final value of the index is a weighted average of the separate measurements. The D$_{n}4000$ index was successfully measured for 381 of the 386 cluster members. The remaining five galaxies only yielded a lower limit to D$_{n}4000$ due to weak or no signal in the blue window of the index, and we did not use them in the following analysis.

At redshift of A963 (and in the local Universe), galaxies form two well-defined sequences of D$_{n}4000$ versus stellar mass \citep{Geller2014, Geller2016, Haines2016,Deshev2017}, one characterised by old stellar populations and a second containing star forming galaxies whose blue light is dominated by young stars. We used the entire ACReS data set ($\sim$ 27000 galaxies), divided into passive and star forming galaxies, according to the absence or presence of emission lines in their spectra, to estimate the mean D$_{n}4000$ for passive galaxies as a function of stellar mass. Figure \ref{D4000hist} shows the distribution of offsets from this relation for all A963 member galaxies split into passive and star forming as described in Sect.\ref{sect_smcgc}, and Fig. \ref{ssfr}. The two samples show two distinct distributions with the minimum between them shown by the grey dotted line. We use this information to define two classes of galaxies which will be analysed in Section \ref{sect_results}. All the passive galaxies to the left of the grey dotted line in Fig.\ref{D4000hist} (higher negative offsets) have D$_{n}4000$ more typical of star forming galaxies and must have had their star formation quenched within the last 0.5 Gyr, $\sim$ 1/3 of the cluster crossing time (assuming $\sigma_{cl}$ = 970 km s$^{-1}$ and R$_{200}$ = 1.85 Mpc, J16). We refer to these galaxies as recently quenched (RQ). All the star forming galaxies that are to the right of the grey line have D$_{n}4000$ more typical of passive galaxies, indicating that the evolved stars dominate their spectral energy distribution (SED). We refer to these galaxies as red star forming (RSF). In total we find 28 RQ and 35 RSF galaxies, constituting 11\% and 27\% of the passive and star forming galaxies, respectively.

The value of D$_{n}4000$ is sensitive to the presence of even a relatively small amount of stars hotter than A-type, which defines the characteristic timescale on which the index reacts to changes in the star formation properties of the galaxy. We used GALEV evolutionary synthesis models \citep{Kotulla2009} and \citet{Bruzual&Charlot2003} spectral evolution models to simulate the evolution of the SED of galaxies depending on their ongoing and past star formation. Passive galaxies with low D$_{n}4000$ values (RQ) are formed only when the truncation of star formation is very rapid or accompanied by a burst of star formation. In the case of high stellar mass galaxies the latter is often required. The modest specific star formation rates typical for galaxies with log(M$_{\star}$) $\geq$ 10 do not usually produce enough massive stars and their integrated light is dominated by the old stellar population. As a consequence their D$_{n}4000$ remains high in the absence of a strong star burst. This issue is further strengthened by the fact that the optical spectroscopy data used here is collected through a single fibre per galaxy. These fibres have limited spatial coverage and are more likely to miss the outskirts of the target galaxies which is often the location with the strongest star formation. A burst producing as little as 1\% of the final stellar mass of the galaxy is sufficient to lower its D$_{n}4000$ to levels more typical of star forming galaxies for a period of a few $\times$ 10$^{8}$ yrs. The rapid quenching and the quenching accompanied by a star burst are expected from galaxies subjected to ram-pressure stripping, where the star bursts occurring in their discs are more centrally concentrated \citep{Jachym2019}. In contrast, galaxies subjected to starvation do not show this decrease in D$_{n}4000$. Their D$_{n}4000$ quickly reaches values typical of passive galaxies even when they still host some residual star formation (like the RSF class). The lifetime of the RQ phase, defined this way, is much longer than the timescales over which other signatures of star formation, such as optical emission lines, disappear from their spectra \citep{KennicutEvans2012}. This allows us to detect galaxies that are likely subjected to ram-pressure stripping and separate them from the ones subjected to starvation. The typical timescales associated with the latter are $\sim$4 Gyr \citep{Peng2015}. The strength of the ram-pressure stripping has strong dependence on the orbit of the galaxy, its orientation, and total mass, among other factors, which define a relatively wide range of timescales on which star formation is expected to be quenched. \citet{TonnesenBryan&vanGorkom2007} estimate between 0.5 and 1.5 Gyr, all significantly faster than starvation, and broadly consistent with the timescales over which D$_{n}4000$ reacts to changes in the star formation rate.

\subsection{21 cm measurements}\label{sect_hidata}
The measurements of the 21 cm emission from atomic hydrogen from galaxies in the field of A963 is based on the data from the BUD{\sc Hi}ES survey (\citealt{Verheijen2007, Deshev2009, Verheijen2010, Jaffe2013}, J16, \citealt{My_thesis}, Gogate et al. \textit{MNRAS accepted}), acquired with the Westerbork Synthesis Radio Telescope (WSRT\footnote[1]{The WSRT is operated by the Netherlands Foundation for Research in Astronomy(NFRA/ASTRON), with financial support by the Netherlands Organisation of Scientific research (NWO)}). The measurements for A963 were collected between 2005 and 2008 and amount to a total of $\sim$ 1400 hours of integration. This was the first of only a handful of blind interferometric {\sc Hi} surveys at such high redshift, and as such represents a unique data set which allows us to estimate the {\sc Hi} content of individual A963 members, and in stacks of spectra over separate parts of the cluster. This estimation was published by J16 as a function of cluster centric distance. In Section \ref{sect_hidef} we present it as a 2D map on the sky.

In order to make the measurements, we extract an {\sc Hi} spectrum at the position and redshift of each star forming cluster member galaxy and stack them within individual Voronoi bins. The extracted spectra spans $\pm$ 850 km s$^{-1}$, from a data cube smoothed in frequency to a resolution of 312.5kHz ($\sim$ 80 km s$^{-1}$). The flux S, integrated over the central $\sim$ 500 km s$^{-1}$ of the stacked spectra is converted to {\sc Hi} mass as M$_{\text{\sc Hi}}$ = 2.36 $\times$ 10$^{5} \times$ d$^{2} \times$ Sd$v$, where d is the distance and d$v$ is the average channel separation, in this case d$v$ = 40 km s$^{-1}$.

\section{Results}\label{sect_results}
   \begin{figure*}
   \centering
   \includegraphics[width=\hsize]{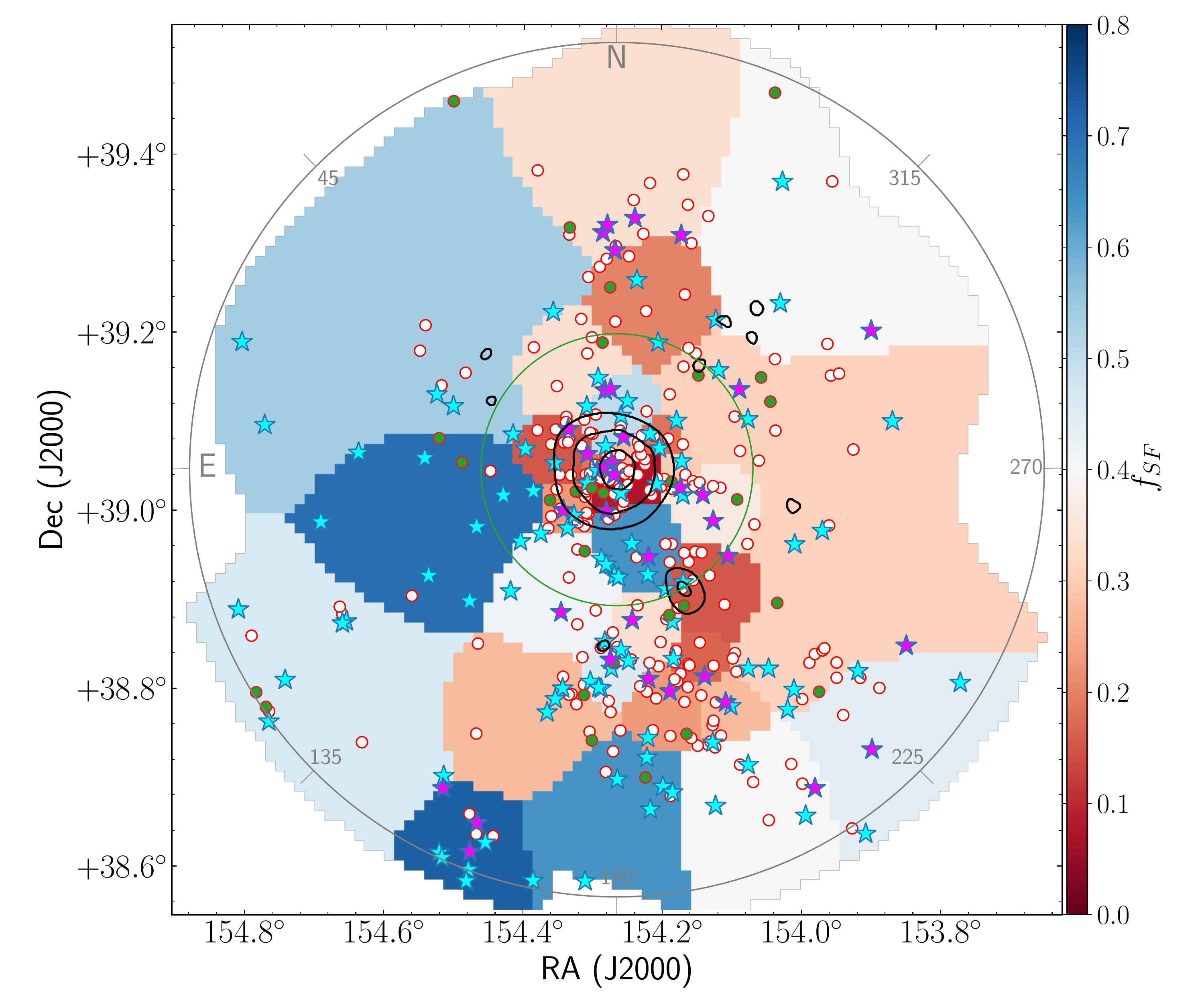}
   \caption{Distribution of A963 member galaxies and their properties on the plane of the sky. The passive galaxies are shown as red bordered circles, with the recently quenched ones (RQ) in green, the rest in white. Star forming galaxies are shown as blue bordered stars, with the anemic ones (RSF) in magenta and the others in cyan. The local fraction of star forming galaxies, $f_{SF}$, mapped on the Voronoi bins is colour-coded according to the colour bar. The green circle in the centre has a radius equal to R$_{200}$ of the cluster. The black contours show the extended X-ray emission. The orientation of the image is shown with the large grey circle, showing the azimuth $\alpha$ as well as the directions towards north and east.}
    \label{fsf}%
    \end{figure*}
   \begin{figure}
   \centering
   \includegraphics[width=\hsize]{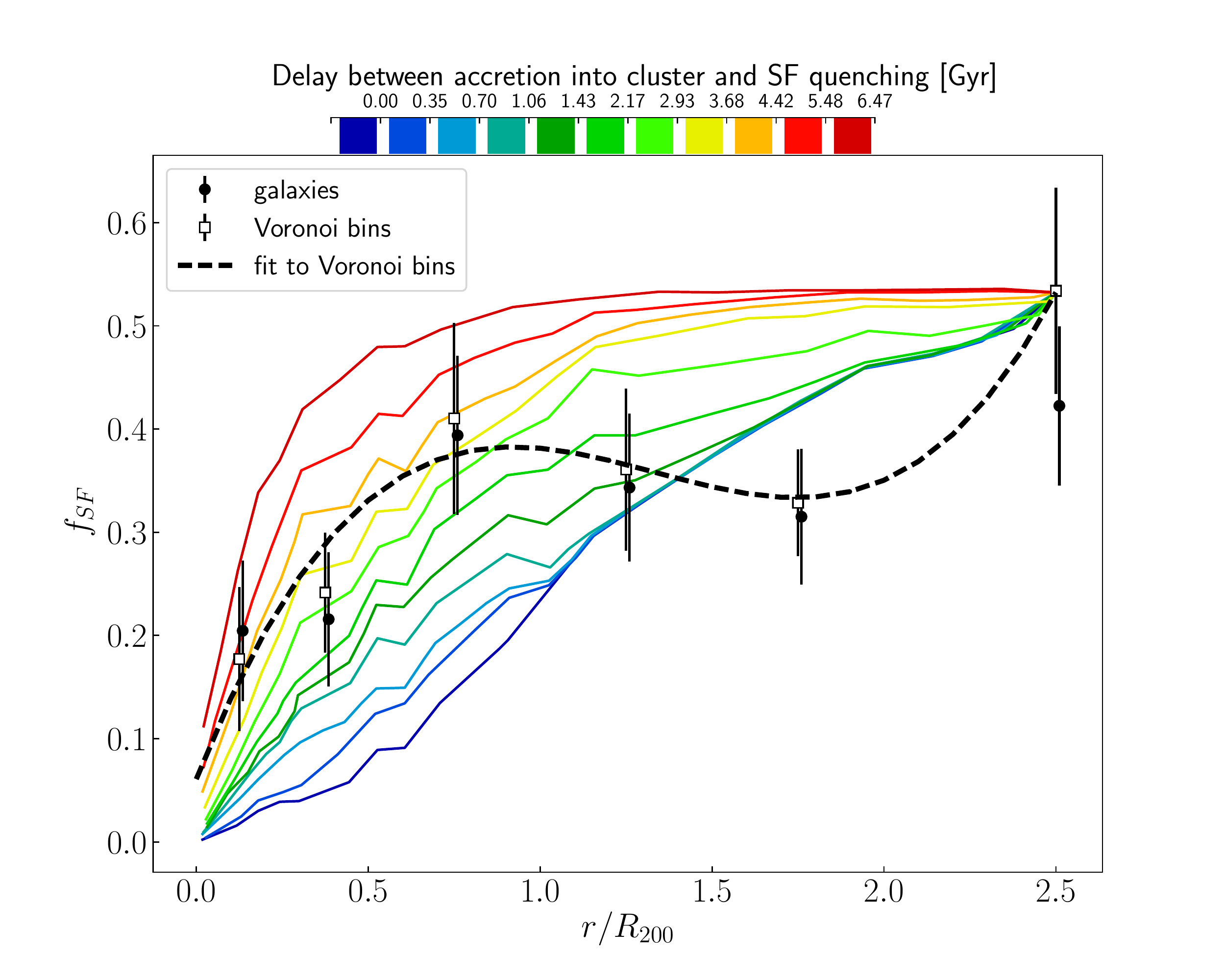}
      \caption{Fraction of star forming galaxies as a function of cluster-centric distance in simulated clusters and in A963. Thin lines of different colour represent galaxies quenched with a delay after their crossing of R$_{200}$ of the parent cluster, according to the colour bar at the top of the figure. The data for A963 are shown in black (symbols and line). The f$_{SF}$ values calculated for individual galaxies are shown as black points, and the mean f$_{SF}$ values within each Voronoi bin are shown as empty squares. The dashed line shows a third-order polynomial fit to the Voronoi bins. For clarity, a small horizontal offset has been added to the black points.}
         \label{model_selection}
   \end{figure}
   \begin{figure*}
   \centering
   \includegraphics[width=\hsize]{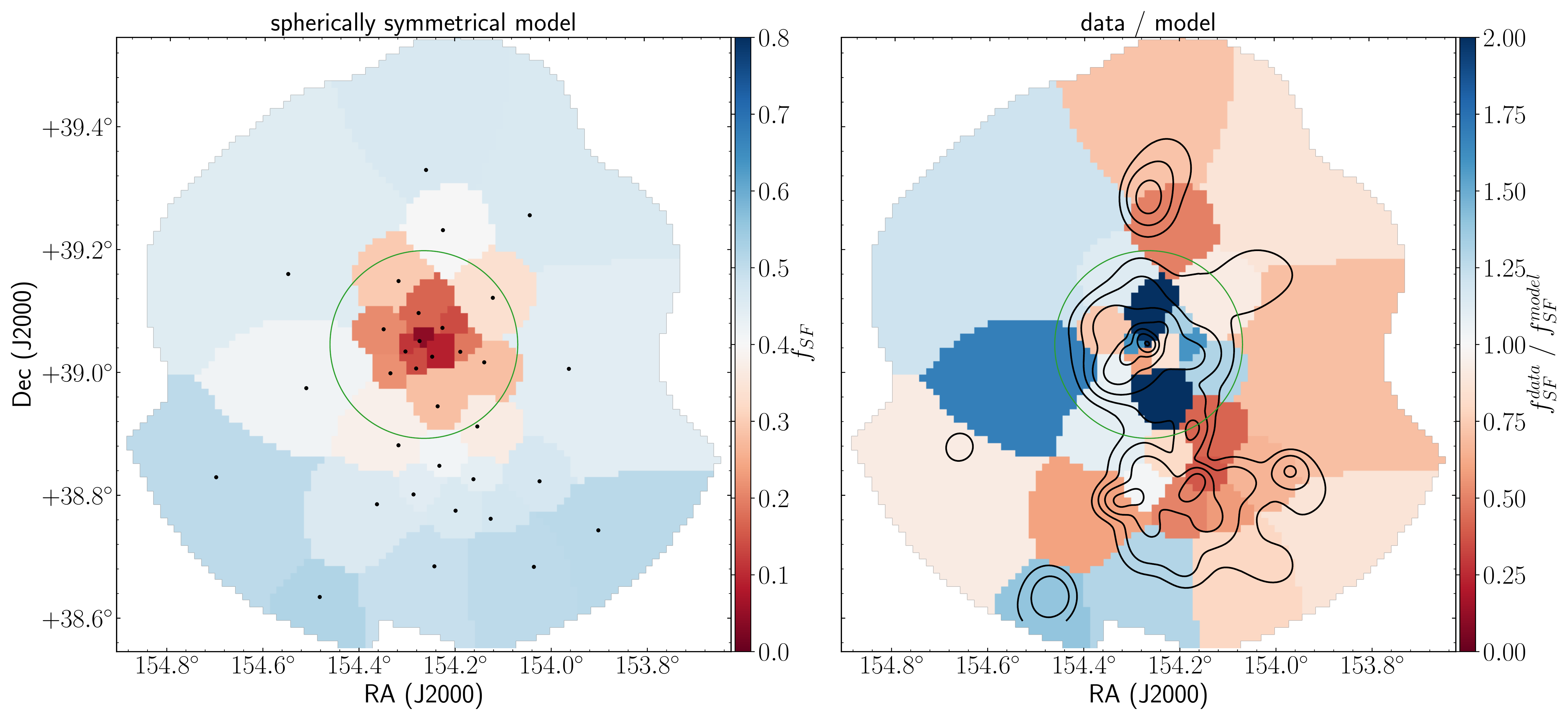}
   \caption{\textit{Left} The model from Fig. \ref{model_selection}, which minimises the difference with the binned data, with a delay of 2.17 Gyr, converted to 2D distribution and used to populate the Voronoi bins with star forming galaxies, according to their K-band luminosity weighted centroids (black points). \textit{Right}- Map of $f_{SF}$, shown in Fig. \ref{fsf}, divided by the spherically symmetrical model shown on the left. Blue bins show an excess of star forming galaxies, red ones show a lack. The black contours show the K-band luminosity density. The lowest contour is at 40 K$^{\star}$ Mpc$^{-2}$. Each following contour is at twice the level of the previous one. The green circle has a radius equal to R$_{200}$ of the cluster.}
    \label{data_model}%
    \end{figure*}
   \begin{figure*}
   \centering
   \includegraphics[width=\hsize]{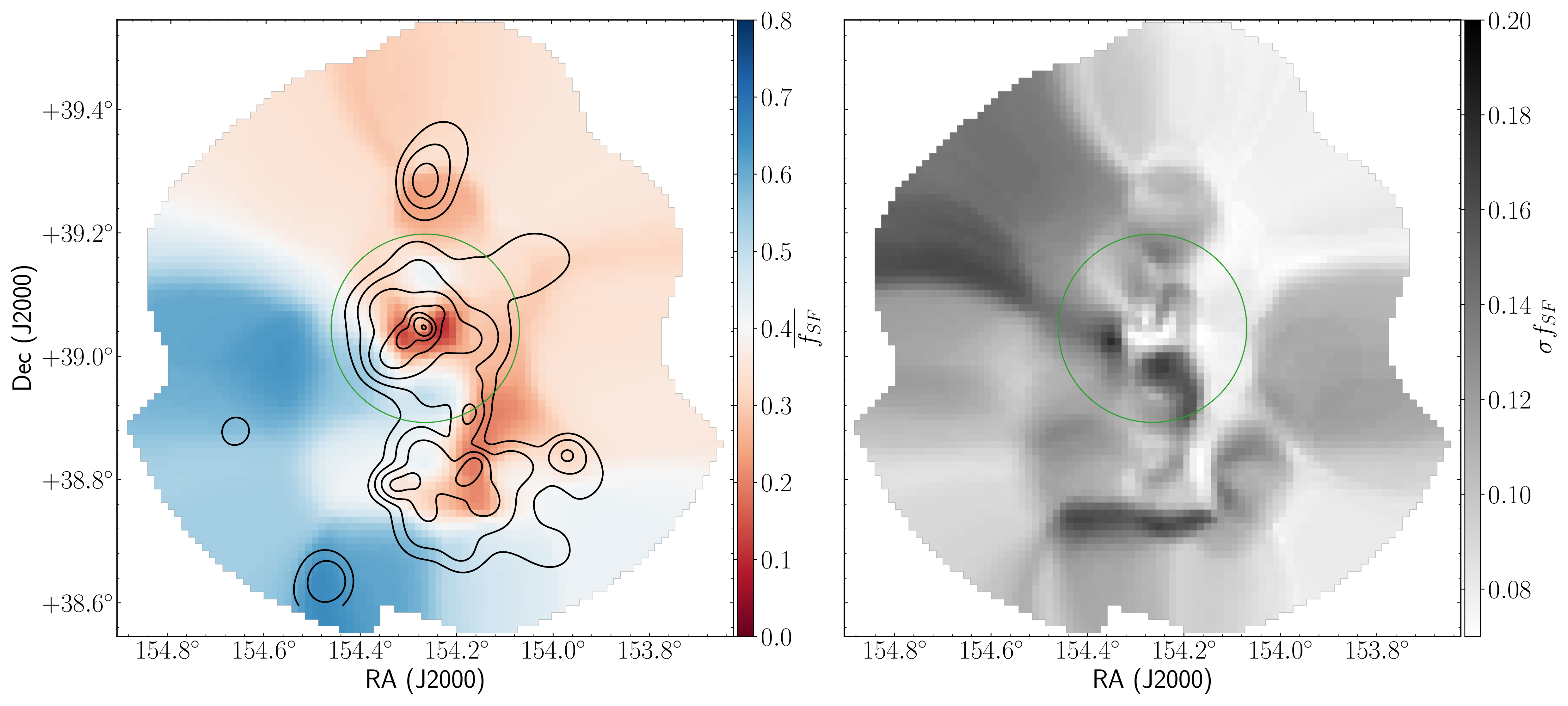}
      \caption{\textit{Left} Map showing the mean fraction of star forming galaxies in Voronoi bins for 1000 jackknife runs on randomly selected 90\% of the data set. \textit{Right} Standard deviation of $f_{SF}$ in these 1000 runs. The black contours show the K-band luminosity density, as in Fig.\ref{data_model}. The green circle has a radius equal to R$_{200}$ of the cluster.}
         \label{jackknife}
   \end{figure*}

\subsection{Distribution of star formation in A963}

In Fig. \ref{fsf} we show a map of the distribution of local fraction of star forming galaxies $f_{SF}$ calculated in Voronoi bins on the sky. The positions of individual galaxies are also indicated with different symbols. Circles with red borders indicate passive galaxies, green for the recently quenched ones (RQ) and white for the rest. Stars with blue borders indicate the star forming galaxies, magenta for the anemic ones (RSF) and cyan for the rest. The green circle in the middle has a radius equal to R$_{200}$ of the cluster. The black contours show the extended X-ray emission from the cluster and other nearby structures, some of which are considered to be groups of galaxies infalling for the first time onto A963 \citep{Haines2018}. The orientation of the image and the azimuthal angle $\alpha$, to which we refer in the following sections, are indicated along the large grey circle. 

In order to quantify the deviation from spherical symmetry of the cluster and its infall region, we employ a model published by \citet{Haines2015}, based on "observations" of the 75 most massive clusters from the Millennium simulations \citep{Springel2005}. The model is shown in Fig. \ref{model_selection}. Galaxies falling onto the average of these 75 clusters have had their star formation quenched instantly at a certain time after their first passing through R$_{200}$ of the cluster. This time delay is shown at the top of the figure, and gives rise to the sequence of thin lines with corresponding colours. The solid black points with error bars show the fraction of star forming galaxies in A963 as a function of cluster centric distance. The same fraction, but measured in Voronoi bins, is shown as empty squares representing the bins with their K-band luminosity weighted centroids. The two sets of points are consistent with each other. The dashed line is a fit of a third-order polynomial to the binned data. The normalisation of the model curves is set so their outermost points have the same $f_{SF}$ as the binned observations at this radius. At r > 1.5 R$_{200}$ the data is not in agreement with the models even within the Poissonian error bars. Inside that radius the data does not follow a trend similar to that of all the models. This is a strong indication that this cluster is not spherically symmetrical in terms of fraction of star forming galaxies. It should be noted that A963 shows a radial trend that is different from the average ACReS cluster, which shows a steady increase in $f_{SF}$ with distance from the centre \citep{Haines2015}. 

By minimising the difference between the range of models and the fit to the measurements in Voronoi bins we find that a delay of 2.17 Gyr is closest to the data, in agreement with \citet{Haines2015}. If only Voronoi bins with r $\leq$ R$_{200}$ were used in the fit a higher delay would be preferred to match the unusually high fraction of star forming galaxies in the centre of this cluster. We use the model with a delay of 2.17 Gyr to populate the Voronoi bins with star forming galaxies, according to the cluster-centric distance of the K-band luminosity weighted centroids of each bin, resulting in the left panel of Fig. \ref{data_model}. The right panel of the figure shows the ratio between the data for A963, presented in Fig. \ref{fsf}, and this spherically symmetrical model. Redder colours represent a lack of star forming galaxies, bluer ones an excess. The model, presented in the left panel of Fig.\ref{data_model}, contains $f_{SF}$ = 0.19 for bins with r $\leq$ R$_{200}$, in perfect agreement with \citet{Butcher&Oemler1984}, who were the first to estimate the unusually high fraction of star forming galaxies hosted by this cluster. The data deviates significantly from the model containing on average, for bins with centroids at r $\leq$ R$_{200}$, 45\% higher $f_{SF}$. However, larger deviations are visible in individual bins. The central bin contains 6 times the expected $f_{SF}$ from the model. To the north and south of it there are bins with 3 and 2.2 times the expected $f_{SF}$. Outside R$_{200}$ the deviations from the model showing lack of star forming galaxies are usually associated with overdensities shown with the black contours. Bin number 16, directly east of the cluster centre, contains ~70\% more star forming galaxies than expected from the model.

There is more than one way of performing adaptive binning of a 2D distribution of points, depending on the distribution of points, the size of the initial grid, and the required population of the Voronoi bins (signal-to-noise ratio). In order to estimate how robust these results are against variations in the sampling of the cluster members and their consecutive binning, we performed 1000 jackknife realisations and present the results in Fig. \ref{jackknife}. For each run we randomly select 90\% of the 386 cluster members and perform the Voronoi binning on this subsample. Then we calculate the fraction of star forming galaxies per bin. The mean and standard deviation of those 1000 maps is shown in the left and right panels of Fig. \ref{jackknife}, respectively. The mean shows qualitatively the same picture as Fig. \ref{fsf}. Because the bins are recalculated every time and because their borders move, the large bin-to-bin variations visible in Fig. \ref{fsf} are smoothed out. This effect is particularly strong closer to the centre of the cluster where the bins are smaller, due to the increased density of galaxies. This presentation of $f_{SF}$ makes the asymmetrical distribution of galaxy properties in A963 more pronounced, and the association of the concentrations of passive galaxies outside R$_{200}$ with overall galaxy concentrations is more clear. Overall, the mapped region outside R$_{200}$ is split into two distinct parts. To the north and west of the cluster, at azimuth 45 < $\alpha$ < 250, the fraction of star forming galaxies is $f_{SF} \simeq$ 0.34. On the opposite side $f_{SF} \simeq$ 0.55 with separate regions at 0.6. Inside R$_{200}$ there is similar bimodality with star forming galaxies dominating the south-eastern part of the cluster as close to the cluster centre as 0.5 R$_{200}$.

We performed tests by varying the size of the initial grid cells and the required S/N in each Voronoi bin and found the results of the jackknife resampling to be very robust against such variations. The right side of Fig.\ref{jackknife} shows the standard deviation of the 1000 jackknife runs. The median standard deviation is 0.1. The real uncertainty is higher as this number does not take into account the uncertainty in estimating star formation rate (median 0.35 dex). We estimate that this variation in the star formation rate would change the overall fraction of star forming galaxies by $\simeq$ 0.1. The uncertainty on the stellar mass (median 0.07 dex) does not play a role since we are separating galaxies using a constant SFR threshold.

We mention one caveat associated with the varying bin size over the image. This makes the resolution of the image also not constant, even though clear variations are visible among neighbouring pixels. Particularly in the outskirts, where individual Voronoi bins change relatively little between the separate jackknife runs, the resolution of the map is much lower than the pixel size.

   \begin{figure*}
   \centering
   \includegraphics[width=\hsize]{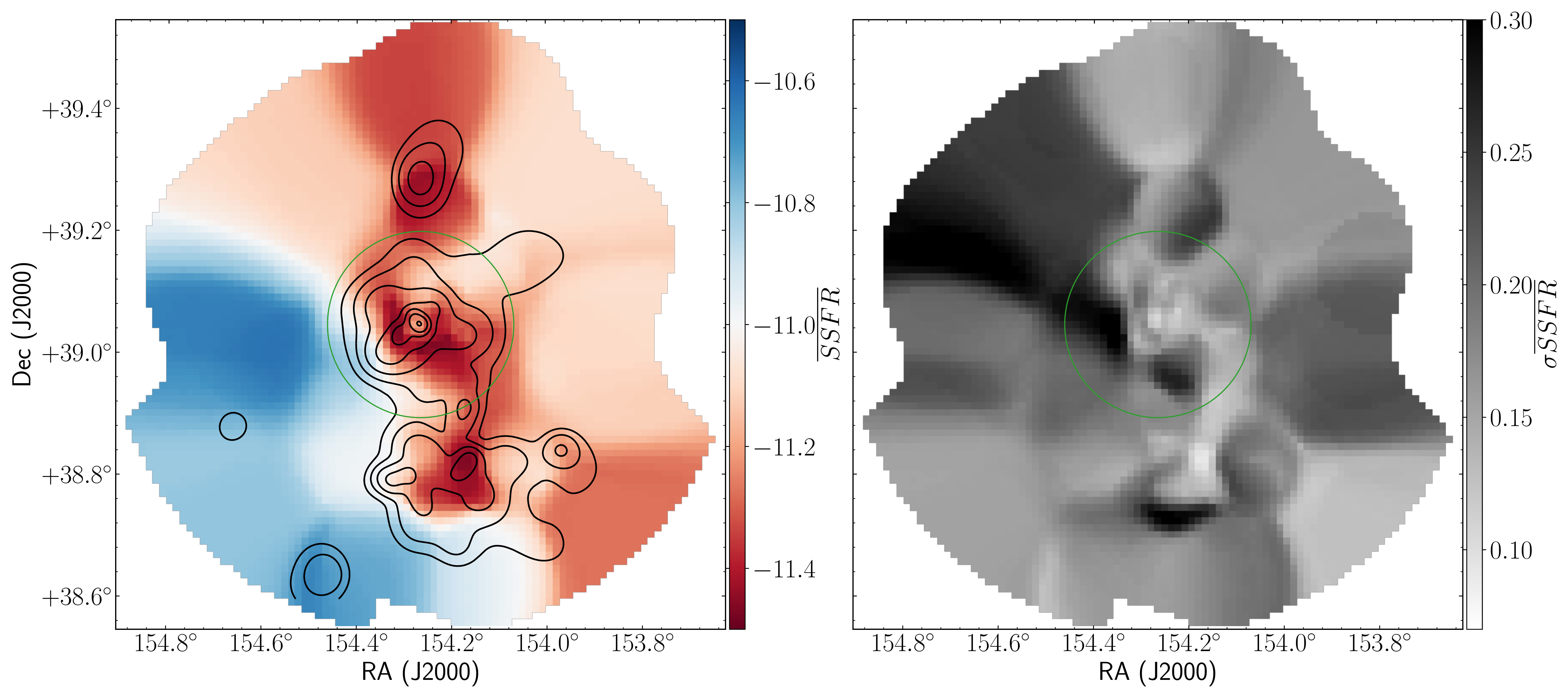}
   \caption{Local mean (\textit{left}) and standard deviation (\textit{right}) of the specific star formation rate in A963 cluster members. The black contours show the K-band luminosity density as in Fig.\ref{data_model}. The green circle has a radius equal to R$_{200}$ of the cluster.}
    \label{SSFR_map}%
    \end{figure*}

Figure \ref{SSFR_map} shows the distribution on the sky of the specific star formation rate in A963. All 386 cluster members were used to produce this map. Qualitatively this map is very similar to the $f_{SF}$ map presented in Fig.\ref{jackknife} but with a few key differences. The same bimodal behaviour is observed outside R$_{200}$ of the cluster, but within it there is a much more pronounced region of reduced SSFR, running north--south through the cluster and extending out to the edges of the map.

   \begin{figure*}
   \centering
   \includegraphics[width=\hsize]{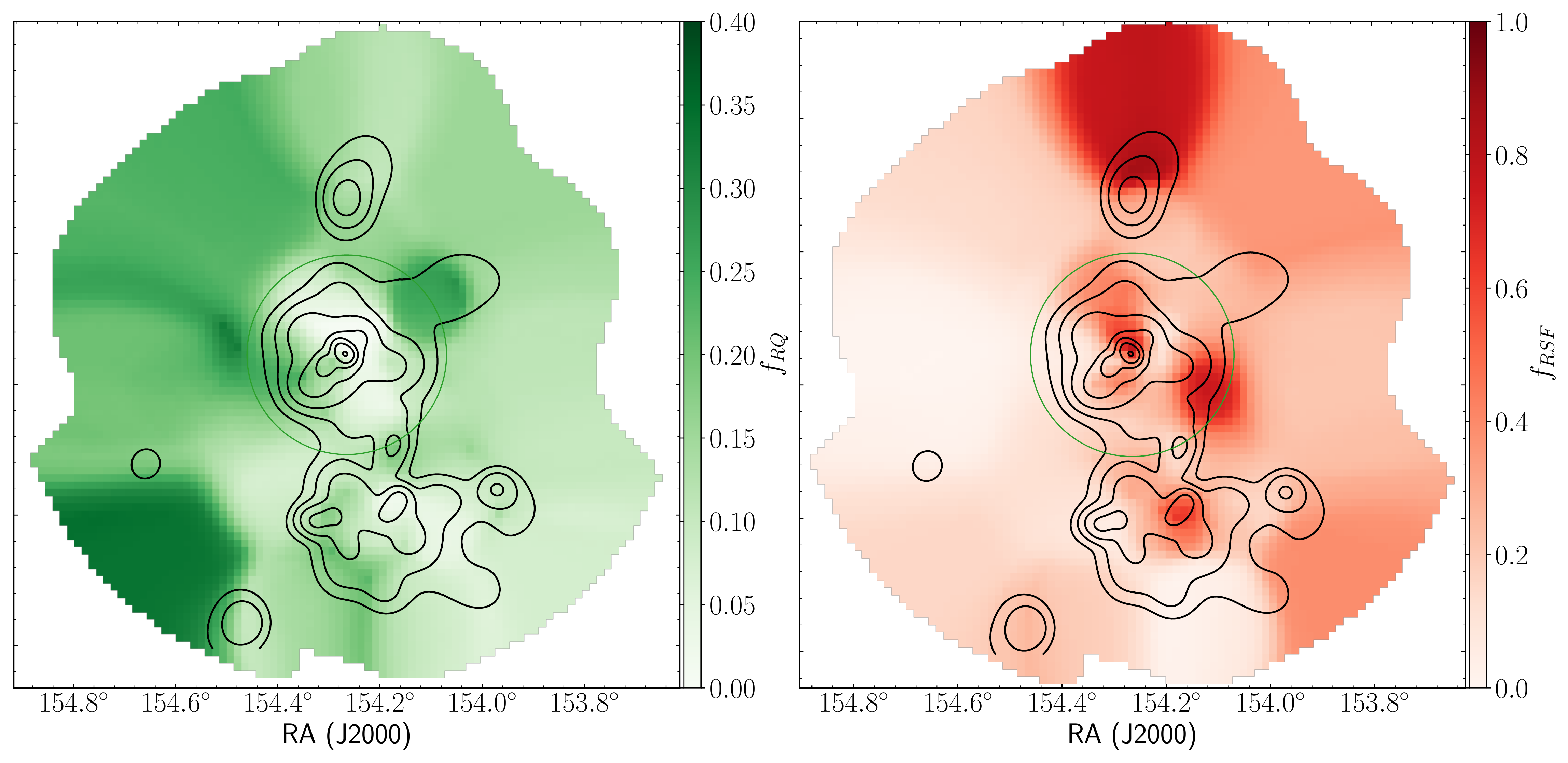}
      \caption{Recently quenched galaxies as a local fraction of the passive galaxies (\textit{left}) and red, star forming galaxies as a local fraction of the star forming galaxies (\textit{right}). The black contours show the K-band luminosity density as in Fig.\ref{data_model}. The green circle has a radius equal to R$_{200}$ of the cluster.}
         \label{RQ_RSF}
   \end{figure*}
\subsection{Recently quenched and red star forming galaxies}\label{sect_RQgals}

The left panel of Fig. \ref{RQ_RSF} shows the distribution on the sky of RQ galaxies as a local fraction of the passive galaxies. As in Figures \ref{jackknife} and \ref{SSFR_map} the mean $f_{RQ}$ from 1000 jackknife runs is shown. The right panel shows the RSF galaxies as a local fraction of the star forming galaxies. The black contours show the luminosity density of cluster members. The two distributions are very different. While the RSF galaxies tend to cluster around the high-luminosity density regions, the RQ galaxies appear to avoid them. The right panel also shows that the concentrations of star forming galaxies inside R$_{200}$ seen in Figures \ref{data_model} and \ref{jackknife} are in fact dominated by RSF, in agreement with Fig.\ref{SSFR_map}. Outside R$_{200}$ the RSF distribution is very similar to the distribution of passive galaxies. There is only one region with elevated luminosity density containing a high fraction of RQ galaxies. The east side of the map is richer in RQ galaxies, and the area directly east of the cluster is where RQ galaxies reach most closely to the cluster centre, as do the star forming galaxies shown in Figures \ref{jackknife} and \ref{SSFR_map}.

\subsection{{\sc Hi} deficiency}\label{sect_hidef}
   \begin{figure*}
   \centering
   \includegraphics[width=\hsize]{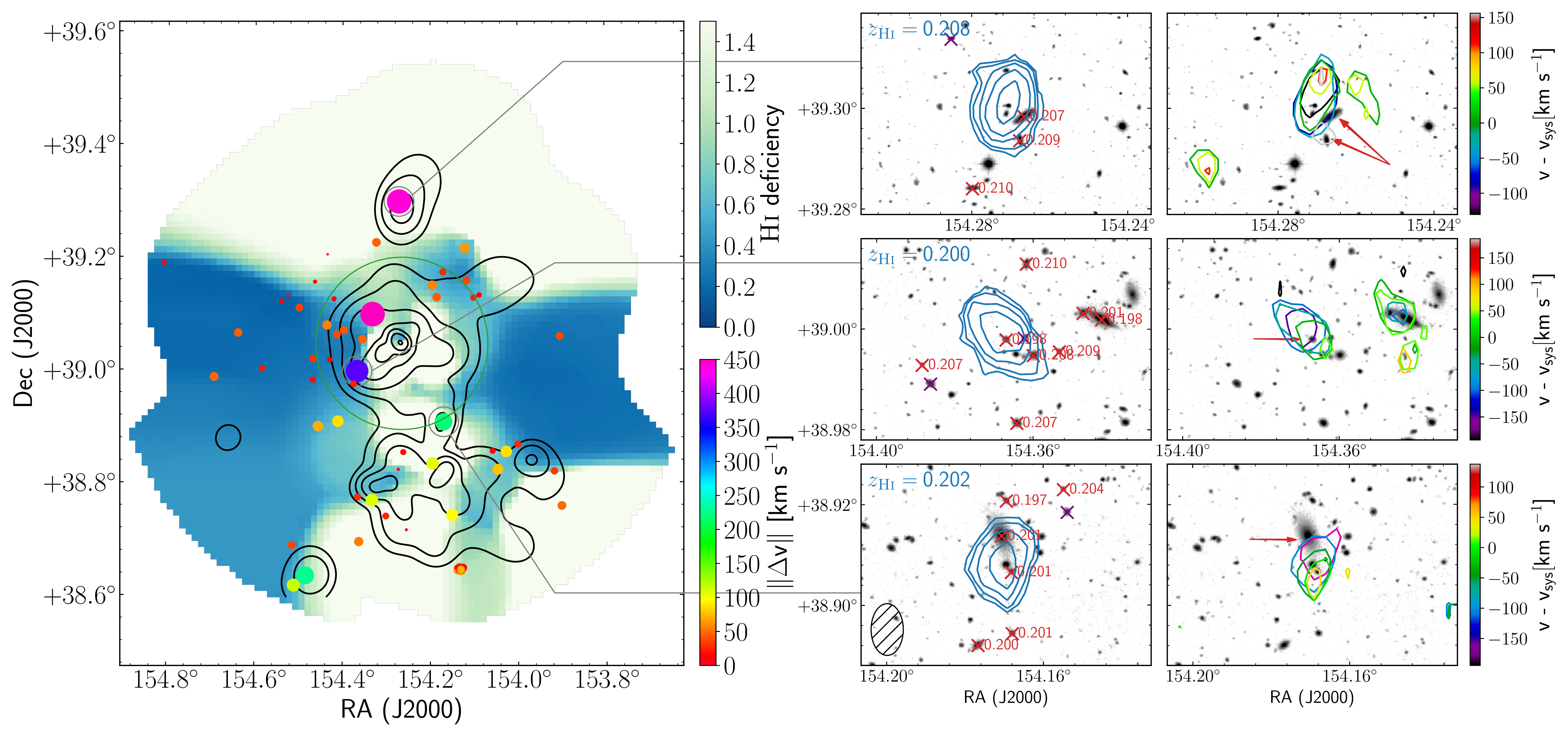}
      \caption{\textit{Left:} Map of the local {\sc Hi} deficiency of star forming galaxies. Lighter colours indicate high deficiency. This is derived from all star forming galaxies. Overlaid are symbols indicating the location of galaxies that are individually detected at 21 cm. The sizes and colours of the symbols are proportional to the offset between the systemic velocities of the gaseous and stellar components. The black contours show the K-band luminosity density, as in Fig.\ref{data_model}. The green circle has a radius equal to R$_{200}$ of the cluster. \textit{Middle:} Total {\sc Hi} maps of three selected sources overplotted on the R-band image. The contours are drawn at 20, 40, 80, 160, and 320 $\mu$Jy beam$^{-1}$. The cluster members with optical spectroscopy are shown in red, and their redshift is indicated. The non-member galaxies with optical spectroscopy are shown in purple. The central redshift of the {\sc Hi} signal is indicated in the top left corner. The beam size is shown in the lowest panel. \textit{Right:} Renzograms of the selected sources. Contours at a constant significance level of 3$\sigma$ are drawn for every channel containing signal from the source. The noise is estimated from the surrounding area of the sources as a function of frequency. The difference between the central velocity of the channels and the central velocity of the entire {\sc Hi} signal is colour-coded. The proposed optical counterparts are indicated with red arrows.}
         \label{HIdef}
   \end{figure*}
The left part of Fig. ref{HIdef} shows a map of the {\sc Hi} deficiency of the star forming galaxies, members of A963. The map is a result of 1000 Voronoi binnings of randomly selected 90\% of the mass complete sample. For every new calculation of the Voronoi bins, we extracted a spectrum from the {\sc Hi} data cube at the location and redshift of all the star forming galaxies in a given bin. These were then brought to the same redshift and averaged. The flux within $\pm$ 250 km s$^{-1}$ of the optical redshift of the galaxy was summed and converted to M$^{observed}_{\texttt{\sc Hi}}$ as explained in Section \ref{sect_hidata}. In order to estimate how {\sc Hi} deficient each galaxy is, this M$^{observed}_{\texttt{\sc Hi}}$ was then compared with the expected {\sc Hi} mass- M$^{expected}_{\texttt{\sc Hi}}$, calculated with a recipe by \citet{Haynes&Giovanelli1984}, based on the optical size of the galaxy, assuming that all the star forming galaxies are of late-type morphology. The optical sizes of A963 galaxies were measured with SExtractor \citep{Bertin1996} as the radius containing 90\% of their light in R band, from the INT mosaic imaging (Section \ref{cluster_membership}). \citet{Haynes&Giovanelli1984} also provide a way of estimating the expected amount of gas for early-type galaxies. We decided not to include the passive galaxies in this analysis because they contain very small amounts of gas and adding them has little effect on the map presented in Fig. \ref{HIdef}. The map shows clearly the central part of A963 as well as the regions to the north and south of the cluster as dominated by galaxies highly deficient in {\sc Hi}. The luminosity density contours trace the regions with high deficiency. From almost all directions, except directly south, less {\sc Hi} deficient galaxies can be found well inside R$_{200}$. Outside this radius the least {\sc Hi} deficient regions are directly east and west of the cluster centre. 

Overlaid on the {\sc Hi} deficiency map are symbols showing the positions of the galaxies detected individually at 21 cm. In total the BUD{\sc Hi}ES survey of A963 detected 127 galaxies (Gogate et al. \textit{MNRAS accepted}). Of these 50 are within the velocity range of the cluster and their proposed optical counterparts are within the stellar mass limits of the sample analysed here. The sizes and colours of individual symbols in Fig.\ref{HIdef} are proportional to the offset in systemic velocity between the {\sc Hi} gas and the stellar component. The combined uncertainty of the two redshift estimates is $<$50 km s$^{-1}$ (in red) and $\sim$ 71\% of the {\sc Hi} detected galaxies are below this value. The remaining high offsets could be the result of confusion, although some show characteristics consistent with the expectation for ram-pressure stripping (discussed in the next section). The confusion is driven by the large size of the synthesised beam of WSRT (shown in the bottom panel of the middle column of Fig.\ref{HIdef}), and by the limited availability of optical spectroscopic data. This prevents us from uniquely identifying the optical counterparts of the {\sc Hi} detections. The plotted offsets are between the central velocity of the {\sc Hi} profile and the optical redshift of the nearest galaxy with available optical redshift. In addition it is very likely that some individual {\sc Hi} detections have multiple optical counterparts.

\subsubsection{Ram-pressure stripping}
A few galaxies were found to have significant offsets in their {\sc Hi} and optical redshift measurements. This difference in velocity may be indicative of ram-pressure stripping. Simulations by \citet{Tonnesen2010} find that in strong ram-pressure stripping, the bulk of the stripped gas is accelerated up to $\sim$60\% of the wind velocity over a period of 250 - 500 Myrs. Given the expected infall velocity of  $\gtrapprox$ 1000 km s$^{-1}$ \citep{Haines2015,Barsanti2016}, this would seem to be a natural explanation for the high velocity-offsets.

The six panels on the right of Fig.\ref{HIdef} shows three galaxies which are among the most plausible candidates for ram-pressure stripping in the A963 cluster (we leave a fuller analysis of the rest of the sample to a future work). As well as the strong velocity offsets, all three show non-circular contours (the renzograms show contours at a constant 3$\sigma$ significance level, colour-coded by velocity channel, following \citet{Rupen1999} and \citet{Kregel2004}. That the contours are non-circular and are found well away from the candidate parent galaxy is strong evidence that the {\sc Hi} is at least partially extended into extragalactic space and not directly associated with optical emission. Simulations by \citet{Taylor2017} have shown that displacing large amounts of {\sc Hi} through tidal encounters, without causing easily visible disturbances in the stellar disc, is extraordinarily difficult, whereas the nature of ram-pressure has no difficulty in removing gas while leaving the stellar disc unaffected. Taken together, this is strongly suggestive of a ram-pressure origin. If so, this would be the first direct imaging of ram-pressure stripping at {\sc Hi} wavelengths at this redshift.

There are two significant caveats to this interpretation. First, due to the large physical size of the WSRT beam at this redshift, we cannot rule out contributions from other galaxies within the beam, which could create the illusion of non-circular contours. Further optical spectroscopic redshift measurements are necessary to fully rule this out, and this scenario is especially plausible for the galaxy in the lower section of figure 13. However, this is much less likely for the other two cases. Given the sensitivity of the {\sc Hi} observations we would expect to only detect the brightest galaxies visible in each image, and while there may be some contribution from the smaller galaxies visible in the upper section, the middle galaxy shows no obvious possible source of confusion.

The second caveat is that, even though it is an indicator of a disturbance, the magnitude of the velocity offset between the {\sc Hi} and optical is much larger than might be expected. Local galaxies have been shown \citep{Oosterloo&vanGorkom2005,Chung2007,Koopmann2008,Leisman2016,Hallenbeck2017,Minchin2019,Taylor2020} to usually have only a small fraction of their {\sc Hi} present in their tails, and retain the bulk of their gas in close proximity to their optical discs. Here, the velocity offsets (200 - 400 km s$^{-1}$) are high enough that the gas may have been entirely displaced from the parent galaxy. While this is highly unusual, it is not completely unknown for local galaxies. For example VCC 1249 in the Virgo cluster has an {\sc Hi} component that is entirely displaced from the stellar emission and a velocity difference of $\sim$ 200 km s$^{-1}$ \citep{Sancisi1987,Henning1993,McNamara1994,Arrigoni2012}. The galaxies described here are much more massive and we therefore expect ram-pressure to be stronger in order to remove their gas, likely resulting in different effects than more typical cases that show only faint {\sc Hi} tails. 
One example of these differences is shown by tuned simulations \citep{Vollmer2001}, which indicate that in massive galaxies the stripped gas can fall back on the stellar disc after the stripping event.
Nonetheless, detailed modelling and additional optical spectroscopy will be necessary to properly confirm the nature of these objects.

\section{Discussion}
\subsection{Connection to the large-scale structure}
   \begin{figure}
   \centering
   \includegraphics[width=\hsize]{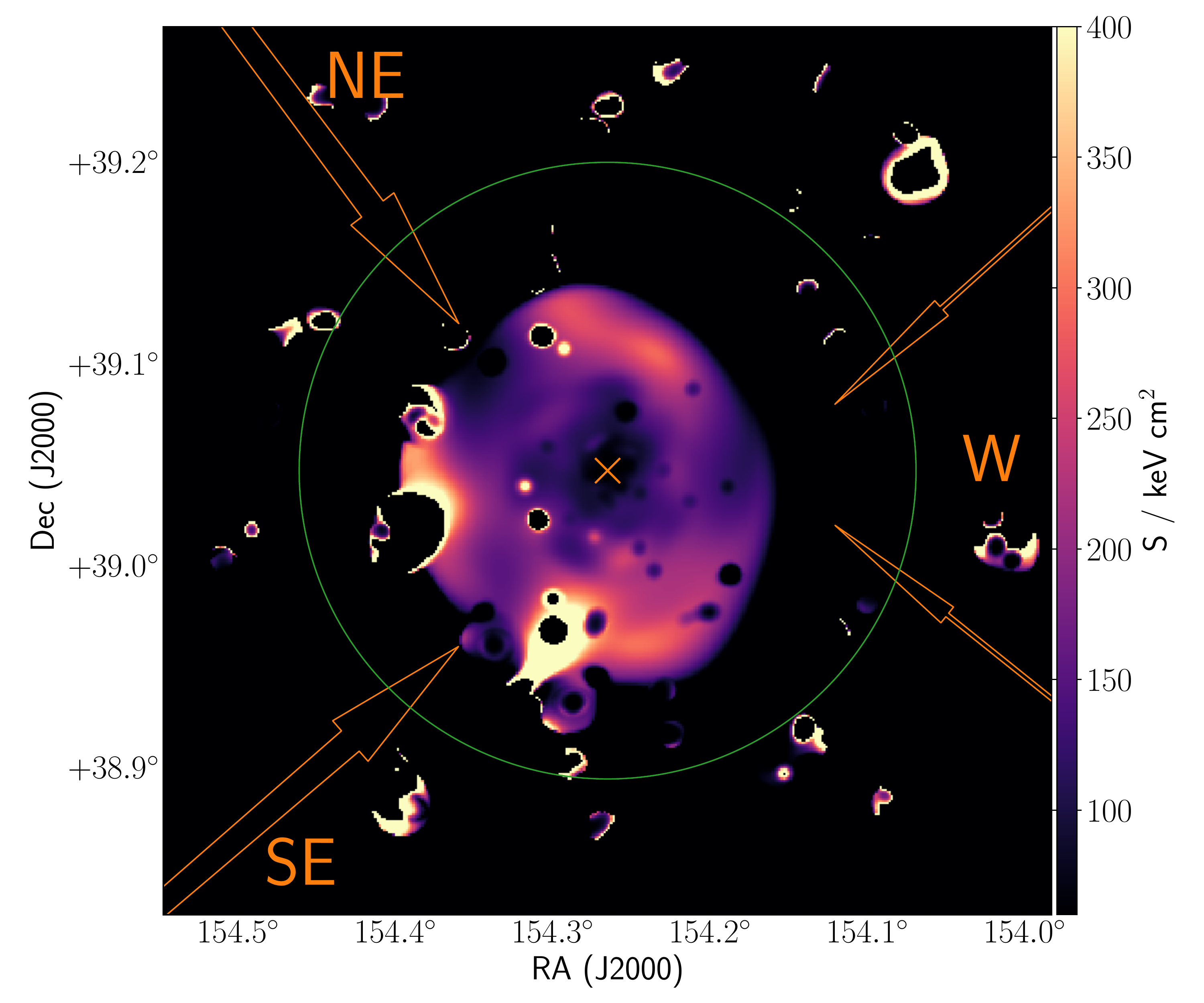}
   \caption{Entropy map of the hot gas of A963, based on XMM-Newton observations. The arrows show the approximate direction of inflow expected from the filaments identified by J16 from the large-scale galaxy distribution around the cluster. The $\times$ symbol in the centre marks the location of the BCG. The green circle has a radius equal to R$_{200}$ of the cluster.}
    \label{entropy}%
    \end{figure}

The results outlined in the previous section suggest a strong dependence of galaxy properties on the direction from which they are infalling onto A963. Naturally we want to trace the infalling galaxies to large cluster-centric distances. However, in the absence of new, more complete spectroscopic data, than used by J16, it is difficult to expand on their conclusions regarding the large-scale structure surrounding A963 and its connectivity. 

\citet{Jaffe2016} proposed the existence of a network of three filaments of the large-scale structure (LSS) crossing at the location of A963 (their Fig.7). Two of them, NE and SE, are relatively tight and well defined; another, W, somewhat more disperse on the sky, although equally coherent in velocity as the first two. This is based on galaxies with available spectroscopic redshift from SDSS. At the redshift of A963, SDSS is complete only for the most massive, usually passive galaxies which, as pointed out by J16, makes the mapping of these filaments tentative at best. The existence and location of these filaments finds a strong support in the entropy map shown in Fig.\ref{entropy}, based on XMM-Newton observations. The figure shows a pseudo-entropy map, obtained using a calibrated hardness ratio map that traces the temperature of the cluster's ICM and the soft band map that traces its density. The arrows show the approximate directions of the proposed three filaments with the W filament bracketed by the two thin arrows. The inflow of low-entropy gas along filaments is observed, at the expected azimuthal angles \citep{Finoguenov2005b}. \cite{Finoguenov2005a} also observed similar low-entropy structure in A3266 likely related to recent accretion of groups and low-mass clusters.

\subsection{Non-spherical cluster}
Clusters of galaxies do not have well-defined shapes as the density distribution of their constituent matter drops progressively as a function of distance to the centre without a well-defined edge \citep{Einasto_profile, NFW}. The distribution of the gravitational mass within the clusters is traceable by strong and weak lensing, but only close to the cluster centre and with considerable uncertainties, those associated with the distribution of background objects, the distribution of their intrinsic orientation, and the presence of other nearby structures along the line of sight \citep{Hwang2014}. Estimating shapes of clusters of galaxies is further complicated by the irregular distribution of the galaxies. The X-ray emitting gas present in the interiors of the clusters can be used as an indicator, but it has a relatively limited radial extent and becomes unobservable at relatively small distances from the centre \citep{Chambers2002}. The distribution of intra-cluster light is known to follow a similar distribution to that of the intra-cluster gas and the overall mass distribution \citep{Montes2019}, but it is also only observable close to the cluster centre. Despite these difficulties the non-sphericity of clusters has long been established observationally \citep{Binggeli1982,Plionis1994,Kasun2005,Donahue2016,Einasto2018a}, and in simulations \citep{Okabe2018}.

We focus on mapping the parts of the clusters where galaxies are being transformed, and so in this section we discuss the shape of A963, as indicated by the varying properties of its member galaxies. This is not necessarily directly related to the density of galaxies (Fig. \ref{data_model}), nor to the distribution of the X-ray emitting gas (Fig. \ref{fsf}). Inside R$_{200}$ the strongly asymmetrical distribution of the fraction of star forming galaxies is clearly visible in Figs. \ref{fsf} and \ref{jackknife}, and is also supported by the distribution of specific SFR (Fig. \ref{SSFR_map}). Directly east of the cluster centre, between R$_{500}$ and R$_{200}$, $\sim$ 50$\pm$10\% of the galaxies are star forming, unlike in any other direction. Within this radial range the fraction of star forming galaxies varies between 10\% and 50\% with a typical uncertainty of $\pm$10\%. Even though the binning does smooth the distribution, the jackknife resampling shows that this result is significant. Apart from the cluster centre, the region to the south-west of it inside R$_{200}$ is the most passive one with only $\sim$ 14\% of the galaxies there showing signs of ongoing star formation. This is likely associated with the infalling group discovered by its X-ray emission (named Grp10 by \citet{Haines2018} and group C by J16) and also visible on the luminosity density plots.

Studies have shown a strong correlation between the shape and orientation of the cluster and those of the brightest cluster galaxy \citep{Fuller1999,Kasun2005}. In addition, the orientation of the cluster is often aligned with the large-scale structure surrounding it \citep{BinneySilk1979, Kasun2005, Faltenbacher2005} and other nearby structures up to 30 Mpc away \citep{Binggeli1982}. In cases where the studied cluster is embedded in a massive supercluster its radio and X-ray emission, BCG and infalling substructures are not only aligned with the cluster itself but also with the supercluster axis \citep{Einasto2018a,Einasto2018b}. We used GALFIT \citep{Peng2002} to model the BCG galaxy of A963 with a double Sersic profile \citep{Sersic1968} finding an axis ratio of the outer isophotes close to 0.5 with a position angle pointing north ($\alpha \simeq 0$). The exact value of $\alpha$ changes with surface brightness from 20 to 340 but remains consistent with the alignment of $\sim$ 20\degr projected on the sky, expected from analysing the orientations of dark matter haloes and their central galaxies in Horizon-AGN cosmological hydrodynamical simulation \citep{Okabe2019}. On the other hand, the X-ray emission shows very small ellipticity. 

\subsection{Anisotropic accretion}

   \begin{figure}
   \centering
   \includegraphics[width=\hsize]{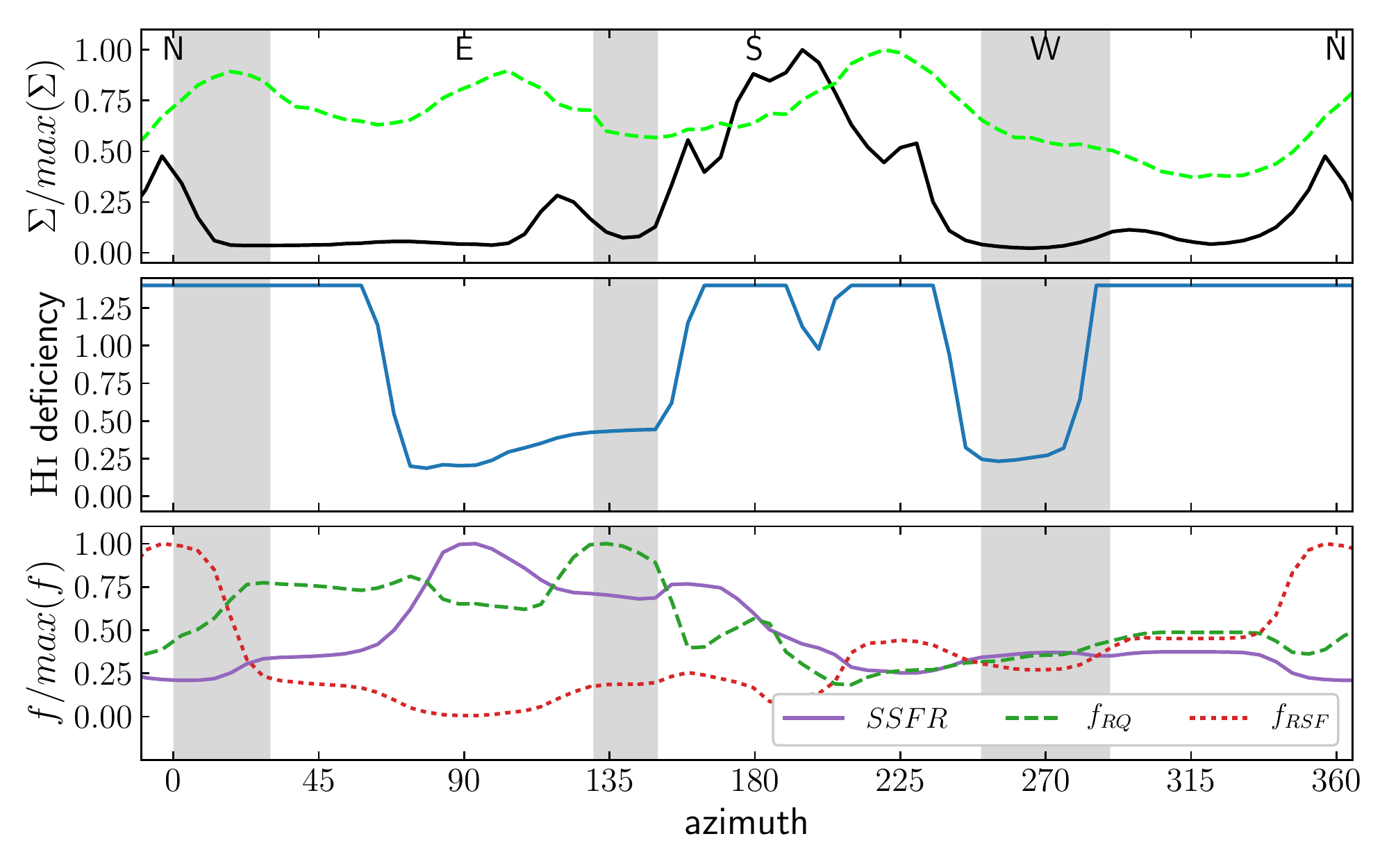}
      \caption{Properties of the galaxy population in the vicinity of A963 and its infall region as a function of the azimuthal angle. The top panel shows the galaxy number density normalised to its maximum. The black curve shows the average K-band luminosity density in the infall region (between 1.5 R$_{200}$ and the edge of the field at $\lesssim$ 3 R$_{200}$). The lime green dashed line shows the mean galaxy number density between 10 and 20 R$_{200}$. The middle panel shows the {\sc Hi} deficiency in the infall region. The bottom panel shows the specific star formation rate (solid purple line), fraction of RQ (dashed green line), and fraction of RSF (red dotted line), expressed as a fraction of their maximum. The grey shaded areas represent the location and relative angular span of the three filaments suggested by J16.}
         \label{azim_props}
   \end{figure}

Before discussing the galaxy properties in the infall region of A963 we note that because of the decreasing galaxy density away from the cluster centre beyond $\sim$ 1.5 R$_{200}$, the binning method we use only provides information as a function of azimuth, with all radial trends masked out by the binning. The only part of the region outside $\sim$ 1.5 R$_{200}$ where we resolve the cluster in the radial direction is at azimuth 170 < $\alpha$ < 225, where the members of the infalling groups visible in Fig.\ref{klum} are raising the galaxy density.

In Fig.\ref{azim_props} we summarise the properties of the galaxies in the infall region of A963 as a function of azimuth. All curves in the figure, except in the middle panel, show the average values over the said radial extent as a function of azimuthal angle, and are normalised to their peak value. The grey shaded regions show the approximate location and azimuth span of the three filaments suggested by J16. The top panel shows the density of galaxies in the infall region of the cluster (1.5 R$_{200} < r \lesssim$ 3 R$_{200}$) with a solid black line, and at 10 R$_{200}$ < r < 20 R$_{200}$ with a lime green dashed line. The former is derived by the K-band luminosity density shown in Fig.\ref{klum}; the latter is a projected number density of SDSS DR12 spectroscopic targets within the velocity range of the cluster. 

 The overall variations in density shown by the lime green line are approximately a factor of two from the peak value with three distinct peaks. One coinciding with the NE filament and the other two located close but not exactly overlapping with the SE and W filaments. The black curve shows much greater variation in density which deviates more from the proposed picture with three filaments.
 
The middle panel shows the {\sc Hi} deficiency in the infall region of the cluster. An artificial saturation value in deficiency is introduced at 1.4 as any values above that are beyond the sensitivity of the survey and are driven by noise. This curve bears very little resemblance to any of the density curves in the top panel and the location of the filaments. The entire northern half of the cluster is devoid of {\sc Hi}. While the southern overdensity is also very deficient, it does contain galaxies with reduced deficiency. The two biggest patches of reduced deficiency while mostly associated with low-density regions also contain higher density regions.

The bottom panel shows the fractions of recently quenched and red star forming galaxies as well as the SSFR of all the galaxies. All the curves are normalised to their peak values. The SSFR peaks, as expected, at the location of the lowest {\sc Hi} deficiency. However, the W filaments which also seem to show decreased {\sc Hi} deficiency does not bring a corresponding increase in SSFR. The overall variation in SSFR is a factor of four. The green dashed curve showing $f_{RQ}$ follows closely the $f_{SF}$ curve, although with somewhat larger variation probably driven by the relatively small number of RQ galaxies. The $f_{RSF}$ curve is almost exactly a negative of the $f_{RQ}$ one, as noted in the discussion of Fig.\ref{RQ_RSF}. The northern part of the cluster seems to offer a very consistent picture of overdensity, traceable out to $\sim$20 R$_{200}$ and populated by gas-poor passive galaxies, few of which have been quenched recently; the galaxies that are not passive host a very low level of star formation. The rest of the infall region is much more difficult to interpret. In particular, the southern overdensity, which at least partially consists of multiple groups (see Fig.\ref{klum}) which contain both passive and star forming galaxies showing a range in {\sc Hi} deficiency, and which have probably arrived at the southern doorstep of the cluster from different directions, considering the presence of three filaments none of which points directly south.

We can try to understand this by comparing with results from simulations, published by \cite{Haines2015}. By observing the 75 most massive clusters in the Millennium simulations, they estimate the fraction of backsplash galaxies in the infall region of these clusters. When defined as the galaxies that have already been inside R$_{200}$ of the cluster, it is expected that $\sim$ 30\% of the galaxies located at 1.5 R$_{200}$ projected from the cluster centre are backsplash galaxies. This fraction drops to 0\% at $\sim$ 3 R$_{200}$. The varying intensity and azimuth of the accretion of matter onto clusters is also a function of time. A similar picture is observed in the ten most massive clusters from Millennium simulations, shown in Fig.9 of \citet{Haines2015}. When considering a single cluster the accretion is likely much more strongly varying in rate and direction with time. In A963 the main accretion happens along the north--south axis. We can expect that the east--west axis would contain a lower fraction of backsplash galaxies, which can partially explain the overdensity of unprocessed galaxies there. Part of the overdensities along the north--south axis is likely due to backsplash galaxies, although many of the individual groups in the south of the cluster are detected in X-ray and by Dressler-Shectman test \citep{DresslerShectman88} presented in J16, which is unlikely for backsplash groups as those would be stripped of their intra-cluster gas and would likely be dispersed during their pericentric passage. It is possible that an accretion of a number of groups at an earlier epoch, along the north--south direction has given a long lasting asymmetric shape of the outskirts of A963. Prolonged anisotropic accretion could mimic the effects of pre-processing, when considering individual clusters, because it will likely lead to strongly asymmetric distribution of backsplash galaxies in the infall region. However, the deviations from the spherically symmetrical model shown in Fig.\ref{data_model} are too large to be entirely due to backsplash galaxies. The northern overdensity is also traceable out to at least 20 $R_{200}$. The presence of red star forming galaxies in this region (Fig. \ref{RQ_RSF}) is a sign that pre-processing is responsible for the quenching of these galaxies, because backsplash galaxies are not expected to have any ongoing star formation. 

While the origin of the passive galaxies situated north of the cluster centre is relatively uncertain, there is little doubt about the gas-rich population located east of the cluster. A963 is an X-ray bright cluster and, as demonstrated by J16, it is unlikely that a galaxy would still host any detectable amount of {\sc Hi} after passing through the pericentre of its orbit. Hence these galaxies must be on their first infall towards the cluster. This is supported by the increase in the fraction of RQ galaxies observed at the same location. The fact that we observe an increased fraction of star forming galaxies well inside R$_{200}$ from the east does not mean they do not experience ram-pressure stripping. We note that the synthesised beam of the WSRT, used for the measurements of the 21 cm emission, is $\sim$ 50 $\times$ 70 kpc at the distance of A963, while the Hectospec fibres used for optical spectroscopy cover the central $\sim$ 5 kpc of the galaxy. This means that in most cases the gas disc of the infalling galaxies can be offset significantly without leaving the WSRT beam. The optical measurements on the other hand only accept light from the central region of the galaxies, which is where the star formation continues the longest before quenching due to ram-pressure stripping. For such galaxies the stripping process could be well under way while we still detect them as unaffected. The two galaxies showing strong velocity offset between the stars and gas in this region support the idea of them being subjected to ram-pressure stripping. We should also add that the initial phase of ram-pressure stripping is likely accompanied by an increase in star formation, particularly in the central parts of the disc \citep{Vulcani2018}.

\section{Conclusions}

We used the optical spectroscopy data from \citet{Haines2013} and \citet{Hwang2014}, together with K- and R-band imaging, X-ray imaging, and the deepest 21 cm blind survey (Gogate et al. \textit{MNRAS accepted}) to try and understand the origin of the galaxies responsible for the strong Butcher-Oemler effect observed in A963, as well as the role played by anisotropic accretion in shaping its galaxy population. We present a series of maps showing the distribution on the sky of galaxy properties in and around A963. The cluster shows a strongly asymmetrical structure both in its interior (r $\leq$ R$_{200}$), and in its infall region (1.5 R$_{200}$ < r < $\sim$ 3 R$_{200}$). These deviations from spherical symmetry cannot be explained by cluster related processes which are expected to show simpler dependence on cluster-centric distance, given the very symmetrical X-ray emission whose peak coincides with the location of the BCG. The reason thus must be related to the large-scale structure surrounding A963 and anisotropic accretion of galaxies both in terms of their number and how pre-processed they are. We find an influx of gas rich, star forming galaxies preferentially from east and west of the cluster centre. They reach deep into the cluster interior, particularly from the east side, and are responsible for the strong Butcher-Oemler effect A963 shows. We also observe signatures of ram-pressure stripping in those areas, and the few passive galaxies there have likely had their star formation quenched recently. The accretion perpendicular to that, from the north and south, is dominated by mostly passive galaxies, often clustered in groups. The few star forming galaxies in these regions show very low specific star formation rates and the passive ones do not show signs of any recent star formation. The connection with the large-scale structure surrounding A963 is, however, inconclusive, being data limited. At cluster-centric distances r < 3 R$_{200}$ the galaxy properties suggest strong accretion along the north--south axis of pre-processed galaxies and a weaker accretion of unprocessed galaxies along the east--west axis. At distances > 3 R$_{200}$  three filaments are visible in the overall galaxy distribution, as suggested by J16. We present strong support for the presence of the three filaments in the form of an entropy map, but the properties of the galaxies they bring vary with the western filament bringing into the cluster less processed galaxies than the NE and SE ones. Alongside this we also present the first, tentative, detection at 21cm of ongoing ram-pressure stripping at $z$ = 0.2.

While this analysis, together with the previously published projected phase-space analysis of A963 \citep[J16]{Jaffe2013}, brings us closer to understanding the processes that define the make-up of a massive, intermediate redshift cluster, many questions still remain unanswered. What is the exact connection of A963 with the large-scale structure surrounding it? Why do areas with similar density in the infall region of the cluster seem to bring in galaxy populations with vastly different star formation properties and gas content? The data of the galaxy distribution on larger scales suggests a fragmentation of the filaments around this cluster; is this real or a limitation of the data? Lastly we would like to stress the importance of considering the full dimensionality and all the information available from astronomical observations, when analysing clusters of galaxies.

\begin{acknowledgements}
We thank the anonymous referee for the numerous suggestions and the swift processing of this article. We thank Pavel J\'{a}chym for numerous fruitful discussions during this work. 
This work was supported by the Czech Science Foundation grant 19-18647S and the institutional project RVO 67985815. This work was also supported by institutional research funding IUT26-2 and IUT40-2  of the Estonian Ministry of Education and Research, by the Centre of Excellence "Dark side of the Universe" (TK133) financed by the European Union through the European Regional Development Fund. This research made use of Astropy,\footnote{http://www.astropy.org} a community-developed core Python package for Astronomy \citep{astropy:2013, astropy:2018}, and the Kapteyn Package for Python \citep{KapteynPackage}. This research made use of NASAs Astrophysics Data System Bibliographic Services. 
\end{acknowledgements}

\bibliographystyle{aa} 
\bibliography{/home/tazio/works/references} 

\begin{thebibliography}{104}
\expandafter\ifx\csname natexlab\endcsname\relax\def\natexlab#1{#1}\fi

\bibitem[{{Arrigoni Battaia} {et~al.}(2012){Arrigoni Battaia}, {Gavazzi},
  {Fumagalli}, {Boselli}, {Boissier}, {Cortese}, {Heinis}, {Ferrarese},
  {C{\^o}t{\'e}}, {Mihos}, {Cuilland re}, {Duc}, {Durrell}, {Gwyn},
  {Jord{\'a}n}, {Liu}, {Peng}, \& {Mei}}]{Arrigoni2012}
{Arrigoni Battaia}, F., {Gavazzi}, G., {Fumagalli}, M., {et~al.} 2012, \aap,
  543, A112

\bibitem[{{Astropy Collaboration} {et~al.}(2013){Astropy Collaboration},
  {Robitaille}, {Tollerud}, {Greenfield}, {Droettboom}, {Bray}, {Aldcroft},
  {Davis}, {Ginsburg}, {Price-Whelan}, {Kerzendorf}, {Conley}, {Crighton},
  {Barbary}, {Muna}, {Ferguson}, {Grollier}, {Parikh}, {Nair}, {Unther},
  {Deil}, {Woillez}, {Conseil}, {Kramer}, {Turner}, {Singer}, {Fox}, {Weaver},
  {Zabalza}, {Edwards}, {Azalee Bostroem}, {Burke}, {Casey}, {Crawford},
  {Dencheva}, {Ely}, {Jenness}, {Labrie}, {Lim}, {Pierfederici}, {Pontzen},
  {Ptak}, {Refsdal}, {Servillat}, \& {Streicher}}]{astropy:2013}
{Astropy Collaboration}, {Robitaille}, T.~P., {Tollerud}, E.~J., {et~al.} 2013,
  \aap, 558, A33

\bibitem[{{Baldwin} {et~al.}(1981){Baldwin}, {Phillips}, \&
  {Terlevich}}]{BPT1981}
{Baldwin}, J.~A., {Phillips}, M.~M., \& {Terlevich}, R. 1981, \pasp, 93, 5

\bibitem[{{Balogh} {et~al.}(1999){Balogh}, {Morris}, {Yee}, {Carlberg}, \&
  {Ellingson}}]{Balogh1999}
{Balogh}, M.~L., {Morris}, S.~L., {Yee}, H.~K.~C., {Carlberg}, R.~G., \&
  {Ellingson}, E. 1999, \apj, 527, 54

\bibitem[{{Balogh} {et~al.}(2000){Balogh}, {Navarro}, \& {Morris}}]{Balogh2000}
{Balogh}, M.~L., {Navarro}, J.~F., \& {Morris}, S.~L. 2000, \apj, 540, 113

\bibitem[{{Barsanti} {et~al.}(2016){Barsanti}, {Girardi}, {Biviano}, {Borgani},
  {Annunziatella}, \& {Nonino}}]{Barsanti2016}
{Barsanti}, S., {Girardi}, M., {Biviano}, A., {et~al.} 2016, \aap, 595, A73

\bibitem[{{Bekki} \& {Couch}(2003)}]{Bekki&Couch2003}
{Bekki}, K. \& {Couch}, W.~J. 2003, \apjl, 596, L13

\bibitem[{{Bekki} {et~al.}(2002){Bekki}, {Couch}, \& {Shioya}}]{Bekki2002}
{Bekki}, K., {Couch}, W.~J., \& {Shioya}, Y. 2002, \apj, 577, 651

\bibitem[{{Bertin} \& {Arnouts}(1996)}]{Bertin1996}
{Bertin}, E. \& {Arnouts}, S. 1996, \aaps, 117, 393

\bibitem[{{Bianconi} {et~al.}(2018){Bianconi}, {Smith}, {Haines}, {McGee},
  {Finoguenov}, \& {Egami}}]{Bianconi2018}
{Bianconi}, M., {Smith}, G.~P., {Haines}, C.~P., {et~al.} 2018, \mnras, 473,
  L79

\bibitem[{{Bianconi} {et~al.}(2020){Bianconi}, {Smith}, {Haines}, {McGee},
  {Finoguenov}, \& {Egami}}]{Bianconi2020}
{Bianconi}, M., {Smith}, G.~P., {Haines}, C.~P., {et~al.} 2020, \mnras, 492,
  4599

\bibitem[{{Binggeli}(1982)}]{Binggeli1982}
{Binggeli}, B. 1982, \aap, 107, 338

\bibitem[{{Binney} \& {Silk}(1979)}]{BinneySilk1979}
{Binney}, J. \& {Silk}, J. 1979, \mnras, 188, 273

\bibitem[{{Boselli} \& {Gavazzi}(2006)}]{Boselli&Gavazzi2006}
{Boselli}, A. \& {Gavazzi}, G. 2006, \pasp, 118, 517

\bibitem[{{Boselli} \& {Gavazzi}(2014)}]{BoselliGavazzi2014}
{Boselli}, A. \& {Gavazzi}, G. 2014, Astronomy and Astrophysics Review, 22, 74

\bibitem[{{Bruzual} \& {Charlot}(2003)}]{Bruzual&Charlot2003}
{Bruzual}, G. \& {Charlot}, S. 2003, \mnras, 344, 1000

\bibitem[{{Bruzual A.}(1983)}]{Bruzual1983}
{Bruzual A.}, G. 1983, \apj, 273, 105

\bibitem[{{Butcher} \& {Oemler}(1984)}]{Butcher&Oemler1984}
{Butcher}, H. \& {Oemler}, Jr., A. 1984, \apj, 285, 426

\bibitem[{{Cappellari} \& {Copin}(2003)}]{Cappellari&Copin2003}
{Cappellari}, M. \& {Copin}, Y. 2003, \mnras, 342, 345

\bibitem[{{Chambers} {et~al.}(2002){Chambers}, {Melott}, \&
  {Miller}}]{Chambers2002}
{Chambers}, S.~W., {Melott}, A.~L., \& {Miller}, C.~J. 2002, \apj, 565, 849

\bibitem[{{Chung} {et~al.}(2007){Chung}, {van Gorkom}, {Kenney}, \&
  {Vollmer}}]{Chung2007}
{Chung}, A., {van Gorkom}, J.~H., {Kenney}, J.~D.~P., \& {Vollmer}, B. 2007,
  \apjl, 659, L115

\bibitem[{{Cowie} \& {Songaila}(1977)}]{Cowie&Songaila1977}
{Cowie}, L.~L. \& {Songaila}, A. 1977, \nat, 266, 501

\bibitem[{{Crone Odekon} {et~al.}(2018){Crone Odekon}, {Hallenbeck}, {Haynes},
  {Koopmann}, {Phi}, \& {Wolfe}}]{Odekon2018}
{Crone Odekon}, M., {Hallenbeck}, G., {Haynes}, M.~P., {et~al.} 2018, \apj,
  852, 142

\bibitem[{{Cybulski} {et~al.}(2016){Cybulski}, {Yun}, {Erickson}, {De la Luz},
  {Narayanan}, {Monta{\~n}a}, {S{\'a}nchez}, {Zavala}, {Zeballos}, {Chung},
  {Fern{\'a}ndez}, {van Gorkom}, {Haines}, {Jaff{\'e}}, {Montero-Casta{\~n}o},
  {Poggianti}, {Verheijen}, {Yoon}, {Deshev}, {Harrington}, {Hughes},
  {Morrison}, {Schloerb}, \& {Velazquez}}]{Cybulski2016}
{Cybulski}, R., {Yun}, M.~S., {Erickson}, N., {et~al.} 2016, \mnras, 459, 3287

\bibitem[{{da Cunha} {et~al.}(2008){da Cunha}, {Charlot}, \&
  {Elbaz}}]{daCunha2008}
{da Cunha}, E., {Charlot}, S., \& {Elbaz}, D. 2008, \mnras, 388, 1595

\bibitem[{{Deshev}(2019)}]{My_thesis}
{Deshev}, B. 2019, PhD thesis, Tartu Observatory, University of Tartu, Estonia,
  2019

\bibitem[{{Deshev} {et~al.}(2017){Deshev}, {Finoguenov}, {Verdugo}, {Ziegler},
  {Park}, {Hwang}, {Haines}, {Kamphuis}, {Tamm}, {Einasto}, {Hwang}, \&
  {Park}}]{Deshev2017}
{Deshev}, B., {Finoguenov}, A., {Verdugo}, M., {et~al.} 2017, \aap, 607, A131

\bibitem[{{Deshev} {et~al.}(2009){Deshev}, {Verheijen}, {van Gorkom},
  {Szomoru}, {Dwarakanath}, {Poggianti}, {Schiminovich}, {Chung}, {Yun}, \&
  {Morrison}}]{Deshev2009}
{Deshev}, B., {Verheijen}, M., {van Gorkom}, J., {et~al.} 2009, in Panoramic
  Radio Astronomy: Wide-field 1-2 GHz Research on Galaxy Evolution, 24

\bibitem[{{Donahue} {et~al.}(2016){Donahue}, {Ettori}, {Rasia}, {Sayers},
  {Zitrin}, {Meneghetti}, {Voit}, {Golwala}, {Czakon}, {Yepes}, {Baldi},
  {Koekemoer}, \& {Postman}}]{Donahue2016}
{Donahue}, M., {Ettori}, S., {Rasia}, E., {et~al.} 2016, \apj, 819, 36

\bibitem[{{Dressler} \& {Shectman}(1988)}]{DresslerShectman88}
{Dressler}, A. \& {Shectman}, S.~A. 1988, \aj, 95, 985

\bibitem[{{Ebeling} \& {Kalita}(2019)}]{Ebeling2019}
{Ebeling}, H. \& {Kalita}, B.~S. 2019, arXiv e-prints, arXiv:1907.12781

\bibitem[{{Einasto}(1965)}]{Einasto_profile}
{Einasto}, J. 1965, Trudy Astrofizicheskogo Instituta Alma-Ata, 5, 87

\bibitem[{{Einasto} {et~al.}(2018{\natexlab{a}}){Einasto}, {Deshev}, {Lietzen},
  {Kipper}, {Tempel}, {Park}, {Gramann}, {Hein{\"a}m{\"a}ki}, {Saar}, \&
  {Einasto}}]{Einasto2018a}
{Einasto}, M., {Deshev}, B., {Lietzen}, H., {et~al.} 2018{\natexlab{a}}, \aap,
  610, A82

\bibitem[{{Einasto} {et~al.}(2018{\natexlab{b}}){Einasto}, {Gramann}, {Park},
  {Kim}, {Deshev}, {Tempel}, {Hein{\"a}m{\"a}ki}, {Lietzen},
  {L{\"a}hteenm{\"a}ki}, {Einasto}, \& {Saar}}]{Einasto2018b}
{Einasto}, M., {Gramann}, M., {Park}, C., {et~al.} 2018{\natexlab{b}}, \aap,
  620, A149

\bibitem[{{Einasto} {et~al.}(2014){Einasto}, {Lietzen}, {Tempel}, {Gramann},
  {Liivam{\"a}gi}, \& {Einasto}}]{Einasto2014}
{Einasto}, M., {Lietzen}, H., {Tempel}, E., {et~al.} 2014, \aap, 562, A87

\bibitem[{{Faltenbacher} {et~al.}(2005){Faltenbacher}, {Allgood},
  {Gottl{\"o}ber}, {Yepes}, \& {Hoffman}}]{Faltenbacher2005}
{Faltenbacher}, A., {Allgood}, B., {Gottl{\"o}ber}, S., {Yepes}, G., \&
  {Hoffman}, Y. 2005, \mnras, 362, 1099

\bibitem[{{Finoguenov} {et~al.}(2005){Finoguenov}, {B{\"o}hringer}, \&
  {Zhang}}]{Finoguenov2005b}
{Finoguenov}, A., {B{\"o}hringer}, H., \& {Zhang}, Y.~Y. 2005, \aap, 442, 827

\bibitem[{{Finoguenov} {et~al.}(2006){Finoguenov}, {Henriksen}, {Miniati},
  {Briel}, \& {Jones}}]{Finoguenov2005a}
{Finoguenov}, A., {Henriksen}, M.~J., {Miniati}, F., {Briel}, U.~G., \&
  {Jones}, C. 2006, \apj, 643, 790

\bibitem[{{Fujita}(2004)}]{Fujita2004}
{Fujita}, Y. 2004, \pasj, 56, 29

\bibitem[{{Fujita} \& {Nagashima}(1999)}]{Fujita&Nagashima1999}
{Fujita}, Y. \& {Nagashima}, M. 1999, \apj, 516, 619

\bibitem[{{Fuller} {et~al.}(1999){Fuller}, {West}, \& {Bridges}}]{Fuller1999}
{Fuller}, T.~M., {West}, M.~J., \& {Bridges}, T.~J. 1999, \apj, 519, 22

\bibitem[{{Geller} {et~al.}(2016){Geller}, {Hwang}, {Dell'Antonio}, {Zahid},
  {Kurtz}, \& {Fabricant}}]{Geller2016}
{Geller}, M.~J., {Hwang}, H.~S., {Dell'Antonio}, I.~P., {et~al.} 2016, \apjs,
  224, 11

\bibitem[{{Geller} {et~al.}(2014){Geller}, {Hwang}, {Fabricant}, {Kurtz},
  {Dell'Antonio}, \& {Zahid}}]{Geller2014}
{Geller}, M.~J., {Hwang}, H.~S., {Fabricant}, D.~G., {et~al.} 2014, \apjs, 213,
  35

\bibitem[{{Goto} {et~al.}(2003){Goto}, {Okamura}, {Sekiguchi}, {Bernardi},
  {Brinkmann}, {G{\'o}mez}, {Harvanek}, {Kleinman}, {Krzesinski}, {Long},
  {Loveday}, {Miller}, {Neilsen}, {Newman}, {Nitta}, {Sheth}, {Snedden}, \&
  {Yamauchi}}]{Goto2003}
{Goto}, T., {Okamura}, S., {Sekiguchi}, M., {et~al.} 2003, \pasj, 55, 757

\bibitem[{{Gunn} \& {Gott}(1972)}]{Gunn&Gott1972}
{Gunn}, J.~E. \& {Gott}, III, J.~R. 1972, \apj, 176, 1

\bibitem[{{Haines} {et~al.}(2018){Haines}, {Finoguenov}, {Smith}, {Babul},
  {Egami}, {Mazzotta}, {Okabe}, {Pereira}, {Bianconi}, {McGee}, {Ziparo},
  {Campusano}, \& {Loyola}}]{Haines2018}
{Haines}, C.~P., {Finoguenov}, A., {Smith}, G.~P., {et~al.} 2018, \mnras, 477,
  4931

\bibitem[{{Haines} {et~al.}(2017){Haines}, {Iovino}, {Krywult}, {Guzzo},
  {Davidzon}, {Bolzonella}, {Garilli}, {Scodeggio}, {Granett}, {de la Torre},
  {De Lucia}, {Abbas}, {Adami}, {Arnouts}, {Bottini}, {Cappi}, {Cucciati},
  {Franzetti}, {Fritz}, {Gargiulo}, {Le Brun}, {Le F{\`e}vre}, {Maccagni},
  {Ma{\l}ek}, {Marulli}, {Moutard}, {Polletta}, {Pollo}, {Tasca}, {Tojeiro},
  {Vergani}, {Zanichelli}, {Zamorani}, {Bel}, {Branchini}, {Coupon}, {Ilbert},
  {Moscardini}, {Peacock}, \& {Siudek}}]{Haines2016}
{Haines}, C.~P., {Iovino}, A., {Krywult}, J., {et~al.} 2017, \aap, 605, A4

\bibitem[{{Haines} {et~al.}(2015){Haines}, {Pereira}, {Smith}, {Egami},
  {Babul}, {Finoguenov}, {Ziparo}, {McGee}, {Rawle}, {Okabe}, \&
  {Moran}}]{Haines2015}
{Haines}, C.~P., {Pereira}, M.~J., {Smith}, G.~P., {et~al.} 2015, \apj, 806,
  101

\bibitem[{{Haines} {et~al.}(2013){Haines}, {Pereira}, {Smith}, {Egami},
  {Sanderson}, {Babul}, {Finoguenov}, {Merluzzi}, {Busarello}, {Rawle}, \&
  {Okabe}}]{Haines2013}
{Haines}, C.~P., {Pereira}, M.~J., {Smith}, G.~P., {et~al.} 2013, \apj, 775,
  126

\bibitem[{{Hallenbeck} {et~al.}(2017){Hallenbeck}, {Koopmann}, {Giovanelli},
  {Haynes}, {Huang}, {Leisman}, \& {Papastergis}}]{Hallenbeck2017}
{Hallenbeck}, G., {Koopmann}, R., {Giovanelli}, R., {et~al.} 2017, \aj, 154, 58

\bibitem[{{Haynes} \& {Giovanelli}(1984)}]{Haynes&Giovanelli1984}
{Haynes}, M.~P. \& {Giovanelli}, R. 1984, \aj, 89, 758

\bibitem[{{Henning} {et~al.}(1993){Henning}, {Sancisi}, \&
  {McNamara}}]{Henning1993}
{Henning}, P.~A., {Sancisi}, R., \& {McNamara}, B.~R. 1993, \aap, 268, 536

\bibitem[{{Hwang} {et~al.}(2014){Hwang}, {Geller}, {Diaferio}, {Rines}, \&
  {Zahid}}]{Hwang2014}
{Hwang}, H.~S., {Geller}, M.~J., {Diaferio}, A., {Rines}, K.~J., \& {Zahid},
  H.~J. 2014, \apj, 797, 106

\bibitem[{{Hwang} \& {Lee}(2009)}]{HwangLee2009}
{Hwang}, H.~S. \& {Lee}, M.~G. 2009, \mnras, 397, 2111

\bibitem[{{Jachym} {et~al.}(2019){Jachym}, {Kenney}, {Sun}, {Combes},
  {Cortese}, {Scott}, {Sivanand am}, {Brinks}, {Roediger}, {Palous}, \&
  {Fumagalli}}]{Jachym2019}
{Jachym}, P., {Kenney}, J. D.~P., {Sun}, M., {et~al.} 2019, arXiv e-prints,
  arXiv:1905.13249

\bibitem[{{Jaff{\'e}} {et~al.}(2011){Jaff{\'e}}, {Arag{\'o}n-Salamanca},
  {Kuntschner}, {Bamford}, {Hoyos}, {De Lucia}, {Halliday}, {Milvang-Jensen},
  {Poggianti}, {Rudnick}, {Saglia}, {Sanchez-Blazquez}, \&
  {Zaritsky}}]{Jaffe2011}
{Jaff{\'e}}, Y.~L., {Arag{\'o}n-Salamanca}, A., {Kuntschner}, H., {et~al.}
  2011, \mnras, 417, 1996

\bibitem[{{Jaff{\'e}} {et~al.}(2018){Jaff{\'e}}, {Poggianti}, {Moretti},
  {Gullieuszik}, {Smith}, {Vulcani}, {Fasano}, {Fritz}, {Tonnesen}, {Bettoni},
  {Hau}, {Biviano}, {Bellhouse}, \& {McGee}}]{Jaffe2018}
{Jaff{\'e}}, Y.~L., {Poggianti}, B.~M., {Moretti}, A., {et~al.} 2018, \mnras,
  476, 4753

\bibitem[{{Jaff{\'e}} {et~al.}(2013){Jaff{\'e}}, {Poggianti}, {Verheijen},
  {Deshev}, \& {van Gorkom}}]{Jaffe2013}
{Jaff{\'e}}, Y.~L., {Poggianti}, B.~M., {Verheijen}, M.~A.~W., {Deshev}, B.~Z.,
  \& {van Gorkom}, J.~H. 2013, \mnras [\eprint[arXiv]{1302.1876}]

\bibitem[{{Jaff{\'e}} {et~al.}(2016){Jaff{\'e}}, {Verheijen}, {Haines}, {Yoon},
  {Cybulski}, {Montero-Casta{\~n}o}, {Smith}, {Chung}, {Deshev},
  {Fern{\'a}ndez}, {van Gorkom}, {Poggianti}, {Yun}, {Finoguenov}, {Smith}, \&
  {Okabe}}]{Jaffe2016}
{Jaff{\'e}}, Y.~L., {Verheijen}, M.~A.~W., {Haines}, C.~P., {et~al.} 2016,
  \mnras, 461, 1202

\bibitem[{{Kasun} \& {Evrard}(2005)}]{Kasun2005}
{Kasun}, S.~F. \& {Evrard}, A.~E. 2005, \apj, 629, 781

\bibitem[{{Kenney} {et~al.}(2004){Kenney}, {van Gorkom}, \&
  {Vollmer}}]{KenneyvanGorkom&Vollmer2004}
{Kenney}, J.~D.~P., {van Gorkom}, J.~H., \& {Vollmer}, B. 2004, \aj, 127, 3361

\bibitem[{{Kennicutt} \& {Evans}(2012)}]{KennicutEvans2012}
{Kennicutt}, R.~C. \& {Evans}, N.~J. 2012, \araa, 50, 531

\bibitem[{{Kewley} {et~al.}(2001){Kewley}, {Dopita}, {Sutherland}, {Heisler},
  \& {Trevena}}]{Kewley2001}
{Kewley}, L.~J., {Dopita}, M.~A., {Sutherland}, R.~S., {Heisler}, C.~A., \&
  {Trevena}, J. 2001, \apj, 556, 121

\bibitem[{{Koopmann} {et~al.}(2008){Koopmann}, {Giovanelli}, {Haynes}, {Kent},
  {Balonek}, {Brosch}, {Higdon}, {Salzer}, \& {Spector}}]{Koopmann2008}
{Koopmann}, R.~A., {Giovanelli}, R., {Haynes}, M.~P., {et~al.} 2008, \apjl,
  682, L85

\bibitem[{{Kotulla} {et~al.}(2009){Kotulla}, {Fritze}, {Weilbacher}, \&
  {Anders}}]{Kotulla2009}
{Kotulla}, R., {Fritze}, U., {Weilbacher}, P., \& {Anders}, P. 2009, \mnras,
  396, 462

\bibitem[{{Kregel} {et~al.}(2004){Kregel}, {van der Kruit}, \& {de
  Blok}}]{Kregel2004}
{Kregel}, M., {van der Kruit}, P.~C., \& {de Blok}, W.~J.~G. 2004, \mnras, 352,
  768

\bibitem[{{Kuutma} {et~al.}(2017){Kuutma}, {Tamm}, \& {Tempel}}]{Kuutma2017}
{Kuutma}, T., {Tamm}, A., \& {Tempel}, E. 2017, \aap, 600, L6

\bibitem[{{Larson} {et~al.}(1980){Larson}, {Tinsley}, \&
  {Caldwell}}]{Larson1980}
{Larson}, R.~B., {Tinsley}, B.~M., \& {Caldwell}, C.~N. 1980, \apj, 237, 692

\bibitem[{{Leisman} {et~al.}(2016){Leisman}, {Haynes}, {Giovanelli},
  {J{\'o}zsa}, {Adams}, \& {Hess}}]{Leisman2016}
{Leisman}, L., {Haynes}, M.~P., {Giovanelli}, R., {et~al.} 2016, \mnras, 463,
  1692

\bibitem[{{Mart{\'\i}nez} {et~al.}(2016){Mart{\'\i}nez}, {Muriel}, \&
  {Coenda}}]{Martinez2016}
{Mart{\'\i}nez}, H.~J., {Muriel}, H., \& {Coenda}, V. 2016, \mnras, 455, 127

\bibitem[{{McNamara} {et~al.}(1994){McNamara}, {Sancisi}, {Henning}, \&
  {Junor}}]{McNamara1994}
{McNamara}, B.~R., {Sancisi}, R., {Henning}, P.~A., \& {Junor}, W. 1994, \aj,
  108, 844

\bibitem[{{Minchin} {et~al.}(2019){Minchin}, {Taylor}, {K{\"o}ppen}, {Davies},
  {van Driel}, \& {Keenan}}]{Minchin2019}
{Minchin}, R.~F., {Taylor}, R., {K{\"o}ppen}, J., {et~al.} 2019, \aj, 158, 121

\bibitem[{{Montes} \& {Trujillo}(2019)}]{Montes2019}
{Montes}, M. \& {Trujillo}, I. 2019, \mnras, 482, 2838

\bibitem[{{Navarro} {et~al.}(1997){Navarro}, {Frenk}, \& {White}}]{NFW}
{Navarro}, J.~F., {Frenk}, C.~S., \& {White}, S. D.~M. 1997, \apj, 490, 493

\bibitem[{{Okabe} {et~al.}(2018){Okabe}, {Nishimichi}, {Oguri}, {Peirani},
  {Kitayama}, {Sasaki}, \& {Suto}}]{Okabe2018}
{Okabe}, T., {Nishimichi}, T., {Oguri}, M., {et~al.} 2018, \mnras, 478, 1141

\bibitem[{{Okabe} {et~al.}(2019){Okabe}, {Nishimichi}, {Oguri}, {Peirani},
  {Kitayama}, {Sasaki}, {Suto}, {Pichon}, \& {Dubois}}]{Okabe2019}
{Okabe}, T., {Nishimichi}, T., {Oguri}, M., {et~al.} 2019, arXiv e-prints,
  arXiv:1911.04653

\bibitem[{{Oosterloo} \& {van Gorkom}(2005)}]{Oosterloo&vanGorkom2005}
{Oosterloo}, T. \& {van Gorkom}, J. 2005, \aap, 437, L19

\bibitem[{{Park} \& {Hwang}(2009)}]{ParkHwang2009}
{Park}, C. \& {Hwang}, H.~S. 2009, \apj, 699, 1595

\bibitem[{{Peng} {et~al.}(2002){Peng}, {Ho}, {Impey}, \& {Rix}}]{Peng2002}
{Peng}, C.~Y., {Ho}, L.~C., {Impey}, C.~D., \& {Rix}, H.-W. 2002, \aj, 124, 266

\bibitem[{{Peng} {et~al.}(2015){Peng}, {Maiolino}, \& {Cochrane}}]{Peng2015}
{Peng}, Y., {Maiolino}, R., \& {Cochrane}, R. 2015, \nat, 521, 192

\bibitem[{{Plionis}(1994)}]{Plionis1994}
{Plionis}, M. 1994, \apjs, 95, 401

\bibitem[{{Poggianti} \& {Barbaro}(1997)}]{PoggiantiBarbaro1997}
{Poggianti}, B.~M. \& {Barbaro}, G. 1997, \aap, 325, 1025

\bibitem[{{Poggianti} {et~al.}(2017){Poggianti}, {Moretti}, {Gullieuszik},
  {Fritz}, {Jaffe}, {Bettoni}, {Fasano}, {Bellhouse}, {Hau}, {Vulcani},
  {Biviano}, {Omizzolo}, {Paccagnella}, {D'Onofrio}, {Cava}, {Sheen}, {Couch},
  \& {Owers}}]{Poggianti2017}
{Poggianti}, B.~M., {Moretti}, A., {Gullieuszik}, M., {et~al.} 2017, ArXiv
  e-prints, arXiv:1704.05086

\bibitem[{{Price-Whelan} {et~al.}(2018){Price-Whelan}, {Sip{\H{o}}cz},
  {G{\"u}nther}, {Lim}, {Crawford}, {Conseil}, {Shupe}, {Craig}, {Dencheva},
  {Ginsburg}, {VanderPlas}, {Bradley}, {P{\'e}rez-Su{\'a}rez}, {de Val-Borro},
  {Paper Contributors}, {Aldcroft}, {Cruz}, {Robitaille}, {Tollerud},
  {Coordination Committee}, {Ardelean}, {Babej}, {Bach}, {Bachetti}, {Bakanov},
  {Bamford}, {Barentsen}, {Barmby}, {Baumbach}, {Berry}, {Biscani}, {Boquien},
  {Bostroem}, {Bouma}, {Brammer}, {Bray}, {Breytenbach}, {Buddelmeijer},
  {Burke}, {Calderone}, {Cano Rodr{\'\i}guez}, {Cara}, {Cardoso}, {Cheedella},
  {Copin}, {Corrales}, {Crichton}, {D{\textquoteright}Avella}, {Deil},
  {Depagne}, {Dietrich}, {Donath}, {Droettboom}, {Earl}, {Erben}, {Fabbro},
  {Ferreira}, {Finethy}, {Fox}, {Garrison}, {Gibbons}, {Goldstein}, {Gommers},
  {Greco}, {Greenfield}, {Groener}, {Grollier}, {Hagen}, {Hirst}, {Homeier},
  {Horton}, {Hosseinzadeh}, {Hu}, {Hunkeler}, {Ivezi{\'c}}, {Jain}, {Jenness},
  {Kanarek}, {Kendrew}, {Kern}, {Kerzendorf}, {Khvalko}, {King}, {Kirkby},
  {Kulkarni}, {Kumar}, {Lee}, {Lenz}, {Littlefair}, {Ma}, {Macleod},
  {Mastropietro}, {McCully}, {Montagnac}, {Morris}, {Mueller}, {Mumford},
  {Muna}, {Murphy}, {Nelson}, {Nguyen}, {Ninan}, {N{\"o}the}, {Ogaz}, {Oh},
  {Parejko}, {Parley}, {Pascual}, {Patil}, {Patil}, {Plunkett}, {Prochaska},
  {Rastogi}, {Reddy Janga}, {Sabater}, {Sakurikar}, {Seifert}, {Sherbert},
  {Sherwood-Taylor}, {Shih}, {Sick}, {Silbiger}, {Singanamalla}, {Singer},
  {Sladen}, {Sooley}, {Sornarajah}, {Streicher}, {Teuben}, {Thomas},
  {Tremblay}, {Turner}, {Terr{\'o}n}, {van Kerkwijk}, {de la Vega}, {Watkins},
  {Weaver}, {Whitmore}, {Woillez}, {Zabalza}, \& {Contributors}}]{astropy:2018}
{Price-Whelan}, A.~M., {Sip{\H{o}}cz}, B.~M., {G{\"u}nther}, H.~M., {et~al.}
  2018, \aj, 156, 123

\bibitem[{{Quilis} {et~al.}(2000){Quilis}, {Moore}, \& {Bower}}]{Quilis2000}
{Quilis}, V., {Moore}, B., \& {Bower}, R. 2000, Science, 288, 1617

\bibitem[{{Rupen}(1999)}]{Rupen1999}
{Rupen}, M.~P. 1999, Astronomical Society of the Pacific Conference Series,
  Vol. 180, {Spectral Line Observing II: Calibration and Analysis}, ed. G.~B.
  {Taylor}, C.~L. {Carilli}, \& R.~A. {Perley}, 229

\bibitem[{{Sancisi} {et~al.}(1987){Sancisi}, {Thonnard}, \&
  {Ekers}}]{Sancisi1987}
{Sancisi}, R., {Thonnard}, N., \& {Ekers}, R.~D. 1987, \apjl, 315, L39

\bibitem[{{Sersic}(1968)}]{Sersic1968}
{Sersic}, J.~L. 1968, {Atlas de Galaxias Australes}

\bibitem[{{Smith} {et~al.}(2010){Smith}, {Khosroshahi}, {Dariush}, {Sanderson},
  {Ponman}, {Stott}, {Haines}, {Egami}, \& {Stark}}]{Smith2010a}
{Smith}, G.~P., {Khosroshahi}, H.~G., {Dariush}, A., {et~al.} 2010, \mnras,
  409, 169

\bibitem[{{Springel} {et~al.}(2005){Springel}, {White}, {Jenkins}, {Frenk},
  {Yoshida}, {Gao}, {Navarro}, {Thacker}, {Croton}, {Helly}, {Peacock}, {Cole},
  {Thomas}, {Couchman}, {Evrard}, {Colberg}, \& {Pearce}}]{Springel2005}
{Springel}, V., {White}, S.~D.~M., {Jenkins}, A., {et~al.} 2005, \nat, 435, 629

\bibitem[{{Stroe} {et~al.}(2015){Stroe}, {Oosterloo}, {R{\"o}ttgering},
  {Sobral}, {van Weeren}, \& {Dawson}}]{Stroe2015a}
{Stroe}, A., {Oosterloo}, T., {R{\"o}ttgering}, H.~J.~A., {et~al.} 2015,
  \mnras, 452, 2731

\bibitem[{{Taylor} {et~al.}(2017){Taylor}, {Davies}, {J{\'a}chym}, {Keenan},
  {Minchin}, {Palou{\v{s}}}, {Smith}, \& {W{\"u}nsch}}]{Taylor2017}
{Taylor}, R., {Davies}, J.~I., {J{\'a}chym}, P., {et~al.} 2017, \mnras, 467,
  3648

\bibitem[{{Taylor} {et~al.}(2020){Taylor}, {K{\"o}ppen}, {J{\'a}chym},
  {Minchin}, {Palou{\v{s}}}, \& {W{\"u}nsch}}]{Taylor2020}
{Taylor}, R., {K{\"o}ppen}, J., {J{\'a}chym}, P., {et~al.} 2020, arXiv
  e-prints, arXiv:2001.03385

\bibitem[{{Tempel} {et~al.}(2014){Tempel}, {Stoica}, {Mart{\'\i}nez},
  {Liivam{\"a}gi}, {Castellan}, \& {Saar}}]{Tempel2014}
{Tempel}, E., {Stoica}, R.~S., {Mart{\'\i}nez}, V.~J., {et~al.} 2014, \mnras,
  438, 3465

\bibitem[{{Terlouw} \& {Vogelaar}(2015)}]{KapteynPackage}
{Terlouw}, J.~P. \& {Vogelaar}, M.~G.~R. 2015, {Kapteyn Package, version 2.3},
  {Kapteyn Astronomical Institute}, Groningen, available from
  \url{http://www.astro.rug.nl/software/kapteyn/}

\bibitem[{{Tonnesen} \& {Bryan}(2010)}]{Tonnesen2010}
{Tonnesen}, S. \& {Bryan}, G.~L. 2010, \apj, 709, 1203

\bibitem[{{Tonnesen} {et~al.}(2007){Tonnesen}, {Bryan}, \& {van
  Gorkom}}]{TonnesenBryan&vanGorkom2007}
{Tonnesen}, S., {Bryan}, G.~L., \& {van Gorkom}, J.~H. 2007, \apj, 671, 1434

\bibitem[{{Verdugo} {et~al.}(2008){Verdugo}, {Ziegler}, \&
  {Gerken}}]{Verdugo2008}
{Verdugo}, M., {Ziegler}, B.~L., \& {Gerken}, B. 2008, \aap, 486, 9

\bibitem[{{Verheijen} {et~al.}(2010){Verheijen}, {Deshev}, {van Gorkom},
  {Poggianti}, {Chung}, {Cybulski}, {Dwarakanath}, {Montero-Castano},
  {Morrison}, {Schiminovich}, {Szomoru}, \& {Yun}}]{Verheijen2010}
{Verheijen}, M., {Deshev}, B., {van Gorkom}, J., {et~al.} 2010, arXiv e-prints,
  arXiv:1009.0279

\bibitem[{{Verheijen} {et~al.}(2007){Verheijen}, {van Gorkom}, {Szomoru},
  {Dwarakanath}, {Poggianti}, \& {Schiminovich}}]{Verheijen2007}
{Verheijen}, M., {van Gorkom}, J.~H., {Szomoru}, A., {et~al.} 2007, \apjl, 668,
  L9

\bibitem[{{Vollmer} {et~al.}(2001){Vollmer}, {Cayatte}, {Balkowski}, \&
  {Duschl}}]{Vollmer2001}
{Vollmer}, B., {Cayatte}, V., {Balkowski}, C., \& {Duschl}, W.~J. 2001, \apj,
  561, 708

\bibitem[{{Vulcani} {et~al.}(2018){Vulcani}, {Poggianti}, {Gullieuszik},
  {Moretti}, {Tonnesen}, {Jaff{\'e}}, {Fritz}, {Fasano}, \&
  {Bettoni}}]{Vulcani2018}
{Vulcani}, B., {Poggianti}, B.~M., {Gullieuszik}, M., {et~al.} 2018, \apjl,
  866, L25

\bibitem[{{Vulcani} {et~al.}(2019){Vulcani}, {Poggianti}, {Moretti},
  {Gullieuszik}, {Fritz}, {Franchetto}, {Fasano}, {Bettoni}, \&
  {Jaff{\'e}}}]{Vulcani2019}
{Vulcani}, B., {Poggianti}, B.~M., {Moretti}, A., {et~al.} 2019, \mnras, 487,
  2278

\bibitem[{{Yoon} {et~al.}(2017){Yoon}, {Chung}, {Smith}, \&
  {Jaff{\'e}}}]{Yoon2017}
{Yoon}, H., {Chung}, A., {Smith}, R., \& {Jaff{\'e}}, Y.~L. 2017, \apj, 838, 81

\end{thebibliography}

\end{document}